\documentclass[sigconf]{acmart}

\usepackage{subcaption}
\usepackage{mathptmx}
\usepackage{amsmath}
\usepackage{multicol}
\usepackage{subfiles}
\usepackage{wrapfig}
\usepackage{array}
\usepackage{amssymb}
\usepackage{tikz}
\usetikzlibrary{arrows,shapes,chains,matrix,positioning,scopes,patterns}
\usepackage[linesnumbered,ruled,vlined]{algorithm2e}
\usepackage{graphicx}
\usetikzlibrary{decorations.pathreplacing}
\usepackage {enumerate}
\graphicspath{{img/}}

\makeatletter
\def\@copyrightspace{\relax}
\makeatother
\renewcommand\footnotetextcopyrightpermission [1]{}

\begin{document}        
\title{Eliminating Tight Coupling using Subscriptions Subgrouping in Structured Overlays}
\author{Muhammad Shafique}
\affiliation {Alumni Carleton University Ottawa Canada}
\email {idresi@protonmail.com}

\begin{abstract}
Advertisements and subscriptions are \textit{tightly coupled} to generate publication routing paths in content--based publish/subscribe systems. Tight coupling requires instantaneous updates in routing tables to generate alternative paths which prevents offering scalable and robust dynamic routing in cyclic overlays when link congestion is detected. We propose, \texttt{OctopiA}, first distributed publish/subscribe system for content--based inter--cluster dynamic routing using purpose--built structured cyclic overlays. \texttt{OctopiA} uses a novel concept of \textit{subscription subgrouping}, which divides subscriptions into disjoint sets called \textit{subscription subgroups}. The purpose--built structured cyclic overlay is divided into identical clusters where subscriptions in each subgroup are broadcast to an exclusive cluster. Our advertisement and subscription forwarding algorithms use subscription subgrouping to eliminate tight coupling to offer \textit{inter--cluster} dynamic routing without requiring updates in routing tables. Experiments on a cluster testbed with real world data show that \texttt{OctopiA} reduces the number of saved advertisements in routing tables by 93\%, subscription broadcast delay by 33\%, static and dynamic publication delivery delays by 25\% and 54\%, respectively. 
\end{abstract}

\maketitle
\raggedbottom

\section{Introduction}
\textit{Content--based Publish/Subscribe (pub/sub)} systems provide asynchronous dissemination of data in event--based distributed applications \cite{MANY_FACES}. A \textit{publisher} publishes the data in form of \textit{publication}, while a \textit{subscriber} registers its interest in form of a \textit{subscription} to receive publications of interest \cite{SIENA_WIDE_AREA}\cite{ PADRESBookChapte}. In \textit{broker--based} pub/sub systems, a dedicated overlay network formed by a set of \textit{brokers}, is used to connect publishers and subscribers anonymously. Pub/sub is an active research area that attracts attention from industry \cite{PNUTS,G_CLOUD_PS,MS_PULSE,KAFKA,WormHole,algo_trading} and academia \cite{work_flow, muthy_thesis, MMOG_Canas, TariqPLEROMA, reza-thesis, JEDI, Rebeca, Kyra, MEDYN, reza-partition, reza-thesis, reza_bulk}. In advertisement--based pub/sub systems, a publisher first broadcasts an advertisement message about the data it wants to publish. The advertisement is saved in \textit{Subscription Routing Table (SRT)} at each broker to form an advertisement--tree rooted at the publisher. To form publication routing paths, a subscription is forwarded in reverse direction to brokers that forwarded advertisements matching that subscription. Subscriptions are saved in \textit{Publication Routing Table (PRT)} \cite{SIENA_WIDE_AREA, PADRESBookChapte} (in rest of this paper, an advertisement--based pub/sub system is referred to as \textit{pub/sub}).

\textit{Tight coupling} indicates that an advertisement and matching subscription have to be saved in routing tables of the \textit{same} broker to generate publication routing paths. A subscription has to be reissued when tree of the matching advertisement changes (or the advertisement is reissued) for some reason. Most pub/sub systems use acyclic topologies which provide a single routing path between a pair of publisher and subscriber. This offers limited flexibility to deal with network condition like link congestion. Multiple paths may be available in a cyclic overlay, only one path at a time can be used for content--based routing with almost no support for dynamic routing \cite{SIENA_WIDE_AREA}. We argue that tight coupling requires instantaneous updates in routing tables to generate new routing paths to avoid congested links in cyclic overlays \cite{MERC}. Updates in routing tables generate extra traffic, cause message loss, and impede in--broker processing thus unsuitable for large networks and delay sensitive applications. Pub/sub systems that use cyclic overlays face additional challenges. (i) API calls create a large number of updates in routing tables because there are no covering techniques available for cyclic overlays --- \textit{subscribe} or \textit{unsubscribe} calls require adding or removing the corresponding subscription in all PRTs, which may cause severe congestion \cite{carz_thesis, Congestion}. The worst impact is expected when a publisher calls \textit{unadvertise} and not only the corresponding advertisement has to be removed from all SRTs but subscription--trees of the matching subscriptions may also be removed or pruned. (ii) An advertisement has to be unique identified to avoid message loops. (iii) Advertisement broadcast generates extra \textit{Inter--broker Message (IMs)} to detect and discard duplicates. (iv) Generating routing paths of minimum lengths is difficult, and to the best of our knowledge, never addressed before (cf. Sec. 2). Indeed, a new architecture is needed to address current research challenges in pub/sub systems that use cyclic overlay.   

We introduce \texttt{OctopiA}, a distributed pub/sub system, which offers inter--cluster dynamic routing by eliminating tight coupling. \texttt{OctopiA} is based on a novel concept called \textit{Subscription Subgrouping} (or \textit{Subgrouping}), which creates disjoint subsets of subscriptions, where each subset is known as \textit{Subscription Subgroup} (or \textit{Subgroup}) (cf. Sec. 5). Subgrouping requires that each subgroup should have distinct subscriptions. For example, for subgroups $\mathsf{S}_{1}$ and $\mathsf{S}_{2}$, there exists no subscription $\mathsf{s}$ such that $\mathsf{s} \in (\mathsf{S}_{1} \cap \mathsf{S}_{2})$. To realize subgrouping, \texttt{OctopiA} leverages a new \textit{Structured Cyclic Overlay Topology (SCOT)} to formulate clusters of brokers where the number of clusters is equal to the number of subgroups. A \textit{cluster--level} subscription broadcast algorithm forwards subscriptions in a subgroup to brokers in a cluster. Advertisements are forwarded to a set of brokers, called a \textit{region}, and create advertisement--trees of length 1 (these patterns of subscription and advertisement forwarding eliminate tight coupling). \texttt{OctopiA} does not assign identifications to advertisements and generates no extra IMs. Since subscriptions in a subgroup are confined to one cluster, \textit{subscribe} or \textit{unsubscribe} calls generate updates in routing tables of brokers of only one cluster without effecting network traffic in other clusters. Similarly, \textit{advertise} or \textit{unadvertise} generate updates in routing tables of only one region without effecting network traffic in other regions. This pattern of forwarding subscription and advertisement improves throughput of a pub/sub in face of concurrent churn events. As subscriptions are confined to only one cluster, \textit{inter--cluster} publication routing may generate \textit{false positives} because some clusters may receive publications but have no interested subscribers. An added novelty in \texttt{OctopiA} is the use of a bit--vector to eliminate \textit{false positives}. 

This paper makes the following contributions. Sec. 2 provides details of background issues in pub/sub systems that use cyclic overlays. Sec. 3 provides a comprehensive overview of the Structured Cyclic Overlay Topology (SCOT) used by \texttt{OctopiA}. Additional structural properties induced in SCOT to offer dynamic routing are discussed in Sec. 4. The novel concept of Subscription Subgrouping is discussed in Sec. 5. Advertisement and subscription broadcast algorithms in clustered SCOT are discussed in Sec. 6 and 7, respectively. Sec. 8 presents our bit--vector (or Cluster Index Vector) mechanism to avoid false positives. Algorithms for static and inter-cluster dynamic routing of publications are discussed in Sec. 9. Evaluation and comparison with identification--based (or TID--based \cite{Li_ADAP}) state--of--the--art is presented in Sec. 10.
\section{Background Issues}
\begin{figure*}
	\tikzstyle{line} = [draw, -latex']
	\def\scaleFac {0.7}
	\begin{subfigure}[b]{0.15\textwidth}
		\captionsetup{width=0.95\linewidth}
		\begin{tikzpicture}
		\path [line, thick] (0, 6) -- (0.7, 6);
		\node[text width=2.3cm] at (2, 6) {\tiny \textit{Adv. trees of P, P1}};
		\path [line, thick, dotted] (0, 5.6) -- (0.7, 5.6);
		\node[text width=2.3cm] at (2, 5.6) {\tiny \textit{Adv. tree of P2}};
		\path [line, thick, dashed] (0, 5.2) -- (0.7, 5.2);
		\node[text width=2.3cm] at (2, 5.2) {\tiny \textit{Sub. tree of S}};
		\draw[color=red, dashed] (0, 4.8) -- (0.7, 4.8);
		\node[text width=2.3cm] at (2, 4.8) {\tiny \textit{Region separator}};
		\draw[color=gray!70, thick] (0, 4.4) -- (0.7, 4.4);
		\node[text width=2.3cm] at (2, 4.4) {\tiny \textit{Overlay link}};
		\draw[line width=0.08cm, color=gray!70] (0, 4) -- (0.7, 4);
		\node[text width=2.3cm] at (2, 4) {\tiny \textit{Overloaded overlay link}};
		\path [line, white] (1, 2.8) -- (2, 2.8);
		\end{tikzpicture}
	\end{subfigure}
	~	
	\begin{subfigure}[b]{0.30\textwidth}
		\captionsetup{width=0.95\linewidth}
		\begin{tikzpicture}
		\def\xInc {1.3}
		\def\x {0}
		\def\y {0}
		\def\yInc {1}
		\tikzstyle{every node} = [thick, scale=1]
		\node[draw, circle, scale=\scaleFac] (1) at (\x+\xInc,\y+\yInc) 		{$1$};
		\node[draw, circle, scale=\scaleFac] (2) at (\x+\xInc*2,\y+\yInc) 		{$2$};
		\node[draw, circle, scale=\scaleFac] (3) at (\x+\xInc*3,\y+\yInc) 		{$3$};
		\node[draw, circle, scale=\scaleFac] (4) at (\x+\xInc,\y) 				{$4$};
		\node[draw, circle, scale=\scaleFac] (5) at (\x+\xInc*2,\y) 			{$5$};
		\node[draw, circle, scale=\scaleFac] (6) at (\x+\xInc*3,\y) 			{$6$};
		\node[draw, rectangle, scale=0.7] (S)	at (\x+\xInc*4-0.4, 0) 			{$S$};
		\node[draw, rectangle, scale=0.7] (P)		 at (\x+0.4, \y+\yInc)		{$P$};
		
		\draw [color=gray!70, thick]  (5) -- (4) (6) -- (5) (3) -- (6) (2) -- (3) (1) -- (2) (4) -- (1) (5) -- (2); 
		\draw [thick] (S) -- (6) (P) -- (1);	
		
		\path [line, thick] (P) to [out=30,in=150] (1);
		\path [line, thick] (1) to [out=300,in=70] (4);
		\path [line, thick] (4) to [out=30,in=155] (5);
		\path [line, thick] (5) to [out=30,in=155] (6);
		\path [line, thick] (1) to [out=30,in=155] (2);
		\path [line, thick] (2) to [out=30,in=155] (3);
		\path [line, dashed, thick] (S) to [out=150,in=30] (6);
		\path [line, dashed, thick] (6) to [out=210,in=330] (5);
		\path [line, dashed, thick] (5) to [out=210,in=330] (4);	
		\path [line, dashed, thick] (4) to [out=120,in=240] (1);
		\draw[color=red, dashed] (\x+1.9,-0.3) -- (\x+1.9,1.8);
		\draw[color=red, dashed] (\x+3.2,-0.3) -- (\x+3.2,1.8);		
		\node[text width=3cm] at (\x+\xInc*2, 1.5) {$R_1$};
		\node[text width=3cm] at (\x+\xInc*3, 1.5) {$R_2$};
		\node[text width=3cm] at (\x+\xInc*4, 1.5) {$R_3$};				
		\end{tikzpicture}
		\caption{\footnotesize Advertisement and subscription broadcast in a cyclic overlay when load on the overlay links is normal.}
		\label{fig:adv1}
	\end{subfigure}
	~ 
	\begin{subfigure}[b]{0.25\textwidth}
		\captionsetup{width=0.95\linewidth}
		\begin{tikzpicture}
		\def\xInc {1.3}
		\def\x {0}
		\def\y {0}
		\def\yInc {1}
		\tikzstyle{every node} = [thick, scale=1]
		\node[draw, circle, scale=\scaleFac] (1) at (\x+\xInc,\y+\yInc) 		{$1$};
		\node[draw, circle, scale=\scaleFac] (2) at (\x+\xInc*2,\y+\yInc) 		{$2$};
		\node[draw, circle, scale=\scaleFac] (3) at (\x+\xInc*3,\y+\yInc) 		{$3$};
		\node[draw, circle, scale=\scaleFac] (4) at (\x+\xInc,\y) 				{$4$};
		\node[draw, circle, scale=\scaleFac] (5) at (\x+\xInc*2,\y) 			{$5$};
		\node[draw, circle, scale=\scaleFac] (6) at (\x+\xInc*3,\y) 			{$6$};
		\node[draw, rectangle, scale=0.8] (S)	at (\x+0.4, 0) 					{$S$};
		\node[draw, rectangle, scale=0.8] (P)		 at (\x+0.4, \y+\yInc)			{$P$};
		
		\draw [color=gray!70, thick]  (5) -- (4) (6) -- (5) (3) -- (6) (2) -- (3) (1) -- (2); 
		\draw [thick] (S) -- (4) (P) -- (1);
		\draw [line width=0.08cm, gray!70] (4) -- (1) (5) -- (2);	
		\path [line, thick] (P) to [out=30,in=150] (1);
		\path [line, thick] (3) to [out=300,in=70] (6);
		\path [line, thick] (5) to [out=210,in=340] (4);
		\path [line, thick] (6) to [out=210,in=340] (5);
		\path [line, thick] (1) to [out=30,in=155] (2);
		\path [line, thick] (2) to [out=30,in=155] (3);
		\path [line, dashed, thick] (S) to [out=30,in=155] (4);
		\path [line, dashed, thick] (4) to [out=30,in=155] (5);
		\path [line, dashed, thick] (5) to [out=30,in=155] (6);	
		\path [line, dashed, thick] (6) to [out=120,in=240] (3);
		\path [line, dashed, thick] (3) to [out=210,in=340] (2);
		\path [line, dashed, thick] (2) to [out=210,in=340] (1);
		\draw[color=red, dashed] (\x+1.9,-0.3) -- (\x+1.9,1.8);
		\draw[color=red, dashed] (\x+3.2,-0.3) -- (\x+3.2,1.8);		
		\node[text width=3cm] at (\x+\xInc*2, 1.5) {$R_1$};
		\node[text width=3cm] at (\x+\xInc*3, 1.5) {$R_2$};
		\node[text width=3cm] at (\x+\xInc*4, 1.5) {$R_3$};				
		\end{tikzpicture}
		\caption{\footnotesize Advertisement and subscription forwarding when the links $l \langle 2,5\rangle$ and $l \langle 3,6\rangle$ are congested.}
		\label{fig:adv2}
	\end{subfigure}
	~
	\begin{subfigure}[b]{0.30\textwidth}
		\captionsetup{width=0.98\linewidth}	
		\begin{tikzpicture}
		\def\xInc {1.3}
		\def\x {0}
		\def\y {0}
		\def\yInc {1}
		\tikzstyle{every node} = [thick, scale=1]
		\node[draw, circle, scale=\scaleFac] (1) at (\x+\xInc,\y+\yInc) 		{$1$};
		\node[draw, circle, scale=\scaleFac] (2) at (\x+\xInc*2,\y+\yInc) 		{$2$};
		\node[draw, circle, scale=\scaleFac] (3) at (\x+\xInc*3,\y+\yInc) 		{$3$};
		\node[draw, circle, scale=\scaleFac] (4) at (\x+\xInc,\y) 				{$4$};
		\node[draw, circle, scale=\scaleFac] (5) at (\x+\xInc*2,\y) 			{$5$};
		\node[draw, circle, scale=\scaleFac] (6) at (\x+\xInc*3,\y) 			{$6$};
		\node[draw, rectangle, scale=0.7] (P2)		 at (\x+0.4, 0) 					{$P2$};
		\node[draw, rectangle, scale=0.7] (P1)		 at (\x+0.4, \y+\yInc)			{$P1$};
		\node[draw, rectangle, scale=0.7] (S)		 at (\x+\xInc*4-0.4, 0)			{$S$};
		
		\draw [color=gray!70, thick]  (5) -- (4) (6) -- (5) (3) -- (6) (2) -- (3) (1) -- (2) (4) -- (1) (5) -- (2); 
		\draw [thick] (P2) -- (4) (P1) -- (1) (S) -- (6);	
		\path [line, thick] (P1) to [out=30,in=150] (1);
		\path [line, thick] (3) to [out=300,in=70] (6);
		\path [line, thick] (5) to [out=210,in=340] (4);
		\path [line, thick] (6) to [out=210,in=340] (5);
		\path [line, thick] (1) to [out=30,in=155] (2);
		\path [line, thick] (2) to [out=30,in=155] (3);
		\path [line, dotted, thick] (P2) to [out=30,in=155] (4);
		\path [line, dotted, thick] (4) to [out=120,in=250] (1);
		\path [line, dotted, thick] (1) to [out=330,in=210] (2);	
		\path [line, dotted, thick] (2) to [out=230,in=120] (5);
		\path [line, dotted, thick] (5) to [out=30,in=155] (6);
		\path [line, dotted, thick] (2) to [out=330,in=210] (3);
		\path [line, dashed, thick] (S) to [out=155,in=25] (6);
		\path [line, dashed, thick] (6) to [out=120,in=240] (3);
		\path [line, dashed, thick] (6) to [out=135,in=45] (5);
		\path [line, dashed, thick] (3) to [out=135,in=45] (2);
		\path [line, dashed, thick] (2) to [out=135,in=45] (1);
		\path [line, dashed, thick] (5) to [out=70,in=290] (2);
		\path [line, dashed, thick] (2) to [out=230,in=310] (1);
		\path [line, dashed, thick] (1) to [out=290,in=70] (4);
		\draw[color=red, dashed] (\x+1.9,-0.3) -- (\x+1.9,1.8);
		\draw[color=red, dashed] (\x+3.2,-0.3) -- (\x+3.2,1.8);		
		\node[text width=3cm] at (\x+\xInc*2, 1.5) {$R_1$};
		\node[text width=3cm] at (\x+\xInc*3, 1.5) {$R_2$};
		\node[text width=3cm] at (\x+\xInc*4, 1.5) {$R_3$};				
		\end{tikzpicture}
		\caption{\footnotesize Overlapping advertisements and subscription forwarding to support dynamic publication routing.}
		\label{fig:adv3}
	\end{subfigure}
	\caption{A cyclic overlay of six brokers is divided into three geographical regions $R_{1}$, $R_{2}$ and $R_{3}$. Each region contains two brokers: ($R_{1}$ contains brokers 1 and 4; $R_{2}$ has brokers 2 and 5; $R_{3}$ has brokers 3 and 6). In (a) and (b), the subscriber S is interested in receiving publications from the publisher P. In (c), S is interested in receiving publications from the publishers P1 and P2.} 
	\label{fig:advsub}
\end{figure*}
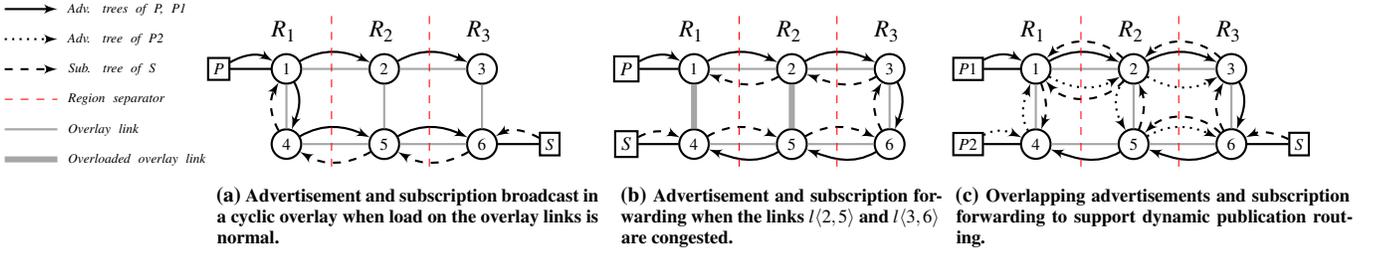

This section describes issues related to handling message loops or cycles in pub/sub systems.\\
\textbf{Advertisement--tree Identification}: 
Cycles generate duplicate advertisements, subscriptions, and publications and it is important to detect and discard them. In the advertisement broadcast, each broker saves the advertisement in the form of \{\textit{advertisement, lasthop}\} tuple where the lasthop indicates the last sender (publisher or broker). Fig. 1(a) shows that Broker 5 may receive the advertisement of P twice from brokers 4 and 2, which generates cycles \cite{Li_ADAP}. To resolve this issue, the host brokers of publishers assign a \textit{Unique Identification (UI)} to each advertisement, which is used by other brokers to detect duplicates \cite{carz_thesis, Li_ADAP, MERC}. In Fig. 1(a), broker 5 discards the advertisement of P when received from broker 2 (assuming that it has already received the advertisement from broker 4). UIs require extra in--broker processing, increase delivery delays and payload of messages \cite{Li_ADAP}.\\
\textbf{Extra Inter--broker Messages (IMs)}: Similar to duplicates in the previous issue in Fig. 1(a), Broker 6 discards the duplicate when received from Broker 3. This indicates that despite using UIs, extra IMs still generate to detect duplicates. The number of these extra IMs could be 80\% of the total network traffic generated in advertisement broadcast \cite{Li_ADAP}. This considerably effects the network performance when churn rate of publishers is high.\\
\textbf{Advertisement--tree Length}: 
In an acyclic overlay, advertisement broadcast generates a unique advertisement--tree even if the advertisement is issued multiple
times (after calling unadvertise from the same broker). Advertisement broadcast in cyclic overlays is an uncontrolled process, which selects the first available link (or broker) as the next destination. This may generate advertisement--trees of larger lengths when load on the links and brokers is uneven. Multiple advertisement calls from the same broker may generate multiple advertisement--trees with different lengths. To the best of our knowledge, no advertisement broadcast algorithm exists that guarantees advertisement--trees of shortest lengths. An overlay link in this paper is represented as $l \langle source, destination \rangle$, where the source identifies message sending broker and the destination represents the message receiving broker. As Fig. 1(b) shows that the length of the advertisement--tree of P is twice of the length in Fig. 1(a). This is due to uneven load on the overlay links $l \langle 1,4 \rangle$ and $l \langle 2,5 \rangle$ at the time P issued the advertisement. Broker 5 receives the advertisement from Broker 6 and discards a duplicate when received from Broker 2. Similarly, Broker 4 discards a duplicate received from Broker 1 (assuming that it has already received the same advertisement from Broker 5). Length of the publication routing path (or the subscription--tree) in this case is also larger than Fig. 1(a). Although P and S are hosted by brokers of $R_{1}$, each publication is processed by brokers in the other two regions to reach Broker 4. This indicates that control over length of the advertisement--tree can reduce publication delay. Hop count to generate shortest advertisement--trees is not a suitable method because it can generate large number of updates in subscription--trees if the advertisement with least hop count is received as duplicate. Ideally, an advertisement--tree should always has the shortest length, even if some links have high loads when the advertisement is issued. \\
\textbf{Binding UI}:
A subscription is bounded to UI of a matching advertisement to avoid cycles in publications routing \cite{Li_ADAP}. A subscription may be bounded to more than one UIs depending on the number of matching advertisements. A broker may receive multiple copies of the same subscription bounded with multiple UIs and coming from different paths. This particular case is explained in Fig. 1(c), where S is interested in publications from P1, and P2. The subscription of S, which is received twice by Broker 2 from the last hops (Broker 3 and Broker 5), bound to UIs of advertisements of P1 and P2. Each publication has to carry UI of the corresponding advertisement--tree for routing to the next destination \cite{Li_ADAP}.\\ 
\textbf{Single Routing Path}:
Only one path \textit{activated} by the advertisement and subscription is available for routing publications, which poses a serious challenge in offering dynamic routing. A naive approach would suggest generating alternative routing paths instantaneously, however, this requires a large number of updates in routing tables to alter advertisement-- and subscription--trees \cite{SIENA_WIDE_AREA, MERC}. In Fig. 1(a), if the link $l \langle 1, 4\rangle$ becomes unavailable, alternative routing paths can be created by removing the link $l \langle 1, 4\rangle$, and adding the link $l \langle 2, 5\rangle$ or $l \langle 3, 6\rangle$ in the advertisement--tree. This requires calling \textit{unadvertise} to eliminate the current advertisement--tree and then generating the desired advertisement--tree using some controlled advertisement broadcast algorithm which is complex for cyclic overlays. In parallel to this, S should be informed to \textit{unsubscribe} and \textit{resubscribe} to receive publications from P. The whole process of removing old and adding new links requires updates in SRTs and PRTs, which is inefficient for large network settings. Dynamic routing is also supported using overlapping advertisement--trees \cite{Li_ADAP}. In Fig. 1(c), advertisement--trees of P1 and P2 overlap at Broker 2, which indicates that Broker 2 can select least loaded link from $l \langle 2, 3\rangle$ and $l \langle 2, 5\rangle$ to forward a publication from P1 or P2 (recall that subscription of S matches with advertisements of P1 and P2). However, this approach requires overlapping advertisement--trees, which are not always possible because subscribers and publishers are anonymous to each other. A dynamically selected path may not have more brokers with overlapping advertisement--trees. In the above example, only one broker can support dynamic routing while a number of brokers have to save the subscription twice \cite{Li_ADAP}.
\section{Structured Cyclic Topology} 
In this section, we describe how Structured Cyclic Overlay Topologies (SCOT) are created using Cartesian Product of Undirected Graphs (CPUG) \cite{theGeneralizedPUG}.
\subsection{Preliminaries}
A graph is an ordered pair $G = (V_{G}, E_{G})$, where $V_{G}$ is a finite set of vertices and $E_{G}$ is a set of edges or links that connect two vertices in $G$. The number of vertices of $G$ (called \textit{order}) is $\mid G\mid$ (or $|V_{G}|$). Similarly, the number of edges in $G$ is $\parallel G \parallel$ (or $|E_{G}|$). A graph in which each pair of vertices are connected by an edge is called a \textit{complete graph}. The diameter of a graph G, represented as \textit{diam(G)}, is the shortest path between the two most distant nodes in $G$.

A graph product is a binary operation that takes two small graph operands---for example $G(V_{G}, E_{G})$ and $H(V_{H}, E_{H})$---to produces a large graph whose vertex set is given by $V_{G} \mathsf{X} V_{H}$. Many types of graph products exist, but we find the Cartesian product most suitable for content-based routing. Other products, for example, the Direct product and the Strong product can be used but their rule-based interconnection of vertices increases node degree and makes routing complex. The $CPUG$ of two graphs $G(V_{G}, E_{G})$ and $H(V_{H}, E_{H})$ is denoted by $G  \square  H$, with vertex set $V_{G \Box H}$ and set of edges $E_{G \Box H}$. Two vertices $(g, h) \in V_{G \Box H}$ and $(g', h') \in V_{G \Box H}$ are adjacent if $g=g'$ and $hh' \in E_{G \Box H}$ or $gg' \in E_{G \Box H}$ and $h=h'$. Formally, the sets of vertices and edges of a CPUG are given as \cite{CPUG_Book}.
\begin{equation}
V_{G \square H} = \{ (g, h) | g \in V_{G}  \wedge  h \in V_{H}\}
\end{equation}
\begin{equation}
\left.\begin{aligned}
E_{G \square H} = \{ \langle (g, h)(g', h') \rangle | (g=g', hh' \in E_{H}) \\ \vee (gg' \in E_{G}, h=h')\}
\end{aligned}
\right\}
\end{equation}
The operand graphs $G$ and $H$ are called factors of $G  \square  H$. CPUG is commutative---that is, $G  \Box  H = H  \square  G$. Although CPUG of $n$ number of graphs is possible, we are concerned with CPUG of only two graphs.
\subsection{Structured Cyclic Overlay Topology}
The \textit{Structured Cyclic Overlay Topology (SCOT)} is a $CPUG$ of two graphs. One graph, represented by $G_{af}$, is called \textit{SCOT acyclic factor}, while the second graph operand, represented by $G_{cf}$, is called \textit{SCOT connectivity factor}. A SCOT has two important properties: (i) \textit{Acyclic Property} emphasizes that the $G_{af}$ must be an acyclic graph, and (ii) \textit{Connectivity Property} requires that $G_{cf}$ must be a complete graph. These properties augment a SCOT with essential characteristics that are used for generating advertisement--trees of shortest lengths and control message cycles efficiently. $V_{af}$ and $V_{cf}$ are the sets of vertices of $G_{af}$ and $G_{cf}$, while $E_{af}$ and $E_{cf}$ are the sets of edges of $G_{af}$ and $G_{cf}$, respectively. For a \textit{singleton graph} of vertex set $\{h\} \subset V_{cf}$, the graph $G_{af}^h$ generated by $G_{af} \Box \{h\}$ is called a $G^h_{af}-fiber$ with \textit{index h}. Similarly, for a singleton graph of vertex set $\{m\} \subset G_{af}$, the graph $G^m_{cf}$ generated by $\{m\} \square G_{cf}$ is called a $G^m_{cf}-fiber$ with \textit{index m}. For each vertex of $G_{cf}$ (and $G_{af}$) $CPUG$ generates one replica of $G_{af}$ (and $G_{cf}$). The number of distinct fibers of $G_{af}$ and $G_{cf}$ is equal to $|V_{cf}|$ and $|V_{af}|$ respectively. We describe the importance of using indexes in SCOT fibers in Sec. 4.
\captionsetup[figure]{labelformat=empty}
\begin{tikzpicture}
\tikzstyle{every node} = [minimum size=7mm]
\def\y {6}
\def\x {1}
\def\xInc {1}
\def\scaleFac {0.7}
\node[draw, thick, circle, scale=\scaleFac] (a) at (\x,\y) {$a$};
\node[draw, thick, circle, scale=\scaleFac] (b) at (\x+\xInc,\y) {$b$};
\node[draw, thick, circle, scale=\scaleFac] (c) at (\x+\xInc*2,\y) {$c$};
\node[draw, thick, circle, scale=\scaleFac] (d) at (\x+\xInc*3,\y) {$d$};
\node[draw, thick, circle, scale=\scaleFac] (e) at (\x+\xInc*4,\y) {$e$};
\node[draw, thick, circle, scale=\scaleFac] (f) at (\x+\xInc*5,\y) {$f$};
\draw [thick] (b) to [out=40,in=140] (d);
\draw [thick] (a) -- (b) (b) -- (c) (d) -- (e) (e) -- (f);
\node[draw, thick, rectangle, scale=0.4] (op)	at (\x+\xInc*6,\y) {};
\node[draw, thick, circle, scale=\scaleFac, fill=green!50] (va) at (\x+\xInc*7,\y) {$1$};
\node[draw, thick, circle, scale=\scaleFac] (vb) at (\x+\xInc*8,\y) {$2$};
\node[draw, thick, circle, scale=\scaleFac, pattern=north west lines, pattern color=gray!40] (vc) at (\x+\xInc*7+0.5,\y+0.7) {$0$};		
\draw [dashed, thick] (va) -- (vb) (vb) -- (vc) (vc) -- (va);
\end{tikzpicture}

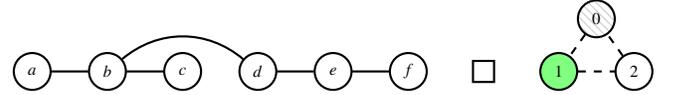
\captionof{figure}{Fig. 2: Operands of CPUG: Left of $\Box$ is $G_{af}$ which is an acyclic graph; right of $\Box$ is $G_{cf}$ which is a complete graph (a triangle).}
\begin{tikzpicture}
\def\scaleSCOT {0.5}
\def\xInc {1.5}
\def\x {0}
\node[draw, thick, circle, scale=\scaleSCOT, pattern=north west lines, pattern color=gray!40] (1) at (\x,6) {$a,0$};
\node[draw, thick, circle, scale=\scaleSCOT, pattern=north west lines, pattern color=gray!40] (2) at (\x+\xInc,6) {$b,0$};
\node[draw, thick, circle, scale=\scaleSCOT, pattern=north west lines, pattern color=gray!40] (3) at (\x+\xInc*2,6) {$c,0$};
\node[draw, thick, circle, scale=\scaleSCOT, pattern=north west lines, pattern color=gray!40] (4) at (\x+\xInc*3,6) {$d,0$};
\node[draw, thick, circle, scale=\scaleSCOT, pattern=north west lines, pattern color=gray!40] (5) at (\x+\xInc*4,6) {$e,0$};
\node[draw, thick, circle, scale=\scaleSCOT, pattern=north west lines, pattern color=gray!40] (6) at (\x+\xInc*5,6) {$f,0$};
\draw [thick] (2) to [out=30,in=150] (4);
\node[draw, thick, circle, scale=\scaleSCOT, fill=green!50] (11) at (\x,5) {$a,1$};
\node[draw, thick, circle, scale=\scaleSCOT, fill=green!50] (12) at (\x+\xInc,5) {$b,1$};
\node[draw, thick, circle, scale=\scaleSCOT, fill=green!50] (13) at (\x+\xInc*2,5) {$c,1$};
\node[draw, thick, circle, scale=\scaleSCOT, fill=green!50] (14) at (\x+\xInc*3,5) {$d,1$};
\node[draw, thick, circle, scale=\scaleSCOT, fill=green!50] (15) at (\x+\xInc*4,5) {$e,1$};
\node[draw, thick, circle, scale=\scaleSCOT, fill=green!50] (16) at (\x+\xInc*5,5) {$f,1$};
\draw [thick] (12) to [out=30,in=150] (14);
\node[draw, thick, circle, scale=\scaleSCOT] (21) at (\x,4) {$a,2$};
\node[draw, thick, circle, scale=\scaleSCOT] (22) at (\x+\xInc,4) {$b,2$};
\node[draw, thick, circle, scale=\scaleSCOT] (23) at (\x+\xInc*2,4) {$c,2$};
\node[draw, thick, circle, scale=\scaleSCOT] (24) at (\x+\xInc*3,4) {$d,2$};
\node[draw, thick, circle, scale=\scaleSCOT] (25) at (\x+\xInc*4,4) {$e,2$};
\node[draw, thick, circle, scale=\scaleSCOT] (26) at (\x+\xInc*5,4) {$f,2$};
\draw [thick] (22) to [out=30,in=150] (24);
\draw [thick] (1) -- (2) (2) -- (3) (4) -- (5) (5) -- (6) (11) -- (12) (12) -- (13) (14) -- (15) (15) -- (16) (21) -- (22) (22) -- (23) (24) -- (25) (25) -- (26);
\draw  [dashed, thick] (1) -- (11) (11) -- (21) (2) -- (12) (12) -- (22) (3) -- (13) (13) -- (23) (4) -- (14) (14) -- (24) 
(5) -- (15) (15) -- (25) (6) -- (16) (16) -- (26);
\draw [dashed, thick] (1) to [out=240,in=120] (21) (2) to [out=240,in=120] (22) (3) to [out=240,in=120] (23) (4) to [out=240,in=120] 
(24) (5) to [out=240,in=120] (25) (6) to [out=240,in=120] (26);
\draw [blue,dotted,thick] (-0.5,6.5) -- (8.3,6.5);
\draw [blue,dotted,thick] (-0.5,5.5) -- (8.3,5.5);
\draw [blue,dotted,thick] (-0.5,4.5) -- (8.3,4.5);
\draw [blue,dotted,thick] (-0.5,3.5) -- (8.3,3.5);
\node[text width=3cm] at (\x+9.5,6) {$C_0$};
\node[text width=3cm] at (\x+9.5,5) {$C_1$};
\node[text width=3cm] at (\x+9.5,4) {$C_2$};
\draw[color=blue, dotted] (\x-0.5,3.5) -- (\x-0.5,6.5);
\draw[color=blue, dotted] (\x+0.7,3.5) -- (\x+0.7,6.5);
\draw[color=blue, dotted] (\x+2.2,3.5) -- (\x+2.2,6.5);	
\draw[color=blue, dotted] (\x+3.7,3.5) -- (\x+3.7,6.5);
\draw[color=blue, dotted] (\x+5.2,3.5) -- (\x+5.2,6.5);	
\draw[color=blue, dotted] (\x+6.7,3.5) -- (\x+6.7,6.5);	
\draw[color=blue, dotted] (\x+7.9,3.5) -- (\x+7.9,6.5);	
\node[text width=2cm] at (\x+0.9,3.3) {$R_a$};
\node[text width=2cm] at (\x+2.4,3.3) {$R_b$};
\node[text width=2cm] at (\x+3.9,3.3) {$R_c$};
\node[text width=2cm] at (\x+5.4,3.3) {$R_d$};
\node[text width=2cm] at (\x+6.9,3.3) {$R_e$};
\node[text width=2cm] at (\x+8.4,3.3) {$R_f$};
\draw (-0.5, 3) -- (0.5, 3);
\node[text width=4cm] at (2.7, 3) {\tiny \textit{Intra-cluster overlay link (aLink)}};
\draw[dashed] (4, 3) -- (5, 3);
\node[text width=4cm] at (7.2, 3) {\tiny \textit{Intet-cluster overlay link (iLink)}};
\end{tikzpicture}
\captionof{figure}{Fig. 3: Structured Cyclic Overlay Topology (SCOT) generated by CPUG of the operands in Fig. 2.\\}
In addition to acyclic and connectivity properties, a SCOT has two more properties: (i) \textit{Index property}, which emphasizes that the labels of nodes of $G_{cf}$ must be a sequence of unique integers starting from zero, and (ii) \textit{Label Order property}, which requires that the first operand (from left to right) of a $CPUG$ node should be from node of $G_{af}$. The index property implies that the index of each fiber of  $G_{af}$ is always an integer. The label order property indicates that the first part of the label of a SCOT node comes from the corresponding vertex of $V_{af}$, and the second part is the label of the corresponding vertex of $V_{cf}$, as indicated by Eq. 1. Reversing the order of operands does not generate extra links or nodes, since CPUG is commutative. These two properties are used for clustering and routing purposes (cf. Sec. 4--9). Fig. 2 shows operands of the SCOT shown in Fig. 3.
\section{Structuredness} 
This section describes a set of classifications for brokers and links to induce a \textit{structuredness} in SCOT. The structuredness divides a SCOT into uniquely identifiable \textit{clusters} and \textit{regions}, which are used in routing advertisements, subscriptions, and publications. 
\subsection{Cluster and Region}
Each $G^i_{af}-fiber$ in a SCOT is a separate group of brokers called a \textit{SCOT Cluster} (or simply a cluster) and represented by $C_{i}$, where $i \in V_{cf}$ is known as \textit{Cluster Index}. A cluster index is the label of a vertex of $V_{cf}$ that generates the cluster (or $G^i_{af}-fiber$) when a CPUG is calculated. Similarly, each $G^j_{cf}-fiber$ is called a \textit{Region} and is represented as $R_{j}$, where $j \in V_{af}$ is a \textit{Region Index}. A region index is the label of a vertex of $V_{cf}$ that generates the region (or $G^j_{cf}-fiber$) when a CPUG is calculated. There are $|V_{cf}|$ and $|V_{af}|$ number of clusters and regions in a SCOT, respectively. SCOT in Fig. 3 contains three clusters (horizontal layers) each identified by $C_{i}$, where $i \in \{0, 1, 2\}$, and six regions (vertical layers) each identified by a unique $R_{j}$, where $j \in {\{a,b,c,d,e,f\}}$. 
\subsection{Classifications}
We assign types to clusters, brokers, and links to classify them for ease in routing advertisements, subscriptions and publications. A SCOT has two types of overlay links: (i) an \textit{intra--cluster overlay link (aLink)}, and (ii) an \textit{inter--cluster overlay link (iLink)}. aLinks connect brokers in the same cluster, while iLinks connect brokers in the same region (Fig. 3). Messaging onto aLinks and iLinks is referred to as \textit{intra--cluster} and \textit{inter--cluster} messaging respectively. There are two types of clusters in a SCOT: (i) a \textit{Primary or Host Cluster}, and (ii) a \textit{Secondary Cluster}. A cluster that contains the host broker of a \textit{client} (publisher or subscriber) is the host or primary cluster, while all other clusters are the secondary clusters of the client. The \textit{Primary neighbours} of a broker are its direct neighbours in the same cluster, while the \textit{Secondary neighbours} (or \textit{Secondary brokers}) are those in the same region. \textit{Host region} of a client is the region that contains host broker of the client. Host cluster of a publisher is its \textit{Target Primary Cluster} (or \textit{Target Cluster}) if at least one matching subscription is available in that cluster. Analogously, a secondary cluster is a \textit{Target Secondary Cluster (TSC) } if at least one matching subscription is available in that cluster. An \textit{edge broker} has at most one primary neighbour, while an \textit{inner broker} has at least two primary neighbours. All brokers in a region are the same type (i.e., are either inner or edge). A SCOT broker in this paper is represented as \textit{B(x, y)}, where $x \in V_{af}$, and $y \in V_{cf}$. In Fig. 4, $C_{0}$ is the host cluster, while $C_{1}$ and $C_{2}$ are secondary clusters of $P_{0}$. Host brokers of $P_{1}$ (i.e, B(e,1)), and $P_{2}$ (i.e, B(a,2)) are inner and edge brokers respectively.
\section{Subscription Subgrouping}
\textit{Subscription Subgrouping (or Subgrouping)} divides subscriptions into multiple disjoint sets, where each subset is known as \textit{Subscription Subgroup (or Subgroup)}. Subgrouping requires that each subgroup should have distinct subscriptions, that is, for subgroups $\mathsf{S}_{1}, \mathsf{S}_{2}, \mathsf{S}_{3}, ..., \mathsf{S}_{n}$; $\mathsf{S}_{1} \cap \mathsf{S}_{2} \cap \mathsf{S}_{3} \cap ... \cap \mathsf{S}_{n} = \emptyset$. Subscriptions in a subgroup are available in the same exclusive overlay network area. This provides multiple benefits: (i) updates in routing tables due to \textit{subscribe} or \textit{unsubscribe} calls effect a limited area of the network, (ii) performance disruption due to any reason can be restricted to a limited network area, (iii) support for community clustering \cite{Com_clustering}, and (v) locality--based QoS provisioning \cite{QoS_Survey}. Importantly, subgrouping could be exploited to eliminate tight coupling to offer dynamic routing without making updates in routing tables. 

\texttt{OctopiA} uses clustered SCOT to realize subgrouping by adopting specific patterns of subscription and advertisement broadcast (Secs. 6 \& 7). The cluster--level subscription broadcast limits the availability of a subscription at a cluster. As envisioned in \cite{QoS_Survey}, QoS management and provisioning in large publish/subscribe systems should be based on locality considerations and could be used to determine partitions of a large overlay network into small logical units, each managing its own independent QoS control and provisioning to have multi--level QoS management. Clusters in a SCOT can be deployed in different geographical locations to connect local and remote publishers with subscribers. Each cluster (and subgroup) can have its own QoS provisioning, thus offering cluster--level demand--based multiple QoS at the same time. Finally, social networking systems like Facebook and Twitter can benefit from subgrouping to adopt \textit{community clustering}, which is known to improve throughput of a pub/sub system by connecting local publishers and subscribers \cite{Com_clustering}. This paper focuses on the use of subgrouping to eliminate tight coupling to offer dynamic routing when congestion is detected.
\section{Advertisement Broadcast}
In \texttt{OctopiA}, the \textit{Advertisement Broadcast Process (ABP)} forwards an advertisement to secondary brokers in the host region of the publisher, while the secondary brokers do not forward the advertisement any further. An advertisement is saved in \textit{Cluster Link Table} (or \textit{CLT}) to create an advertisement--tree of length 1, which is used for inter--cluster routing of publications (cf. Sec. 9). In Fig. 4, the advertisements issued by $P_{0}$, $P_{1}$, and $P_{2}$ are broadcast to brokers in their respective host regions, which makes each advertisement available at most one broker of each cluster. In parallel to help in eliminating tight coupling, this approach has some additional benefits: (i) size of routing table reduced dramatically, (ii) updates in CLT generate only in host region of a publisher when \textit{advertise} or \textit{unadvertise} are called, and (iii) an \textit{unadvertise} call does not require any possible \textit{unsubscribe} calls. \texttt{OctopiA} assigns no identification to advertisement--trees and generates no IMs to detect and discard duplicates. When an \textit{unadvertise} call is received, \texttt{OctopiA} simply removes the advertisement from CLTs of brokers in the host region.
\begin{tikzpicture}
\def\scaleSCOT {0.5}
\def\scaleBox {0.7}
\tikzstyle{line} = [draw, -latex']
\tikzstyle{lineR} = [draw, latex-']
\def\y {6}
\def\x {2.3}
\def\xInc {0.9}
\def\yInc {1.1}
\node[draw, circle, scale=1] (lbroker) at (-1,6) 			  {};
\node[text width=1.5cm] at (0, 6) {\scriptsize \textit{Broker}};
\node[draw, rectangle, scale=1] (lpub)	at  (-1, 5.6) 		{};
\node[text width=1.5cm] at (0, 5.6) {\scriptsize \textit{Publisher}};		
\path [line, dashed, thick] (-1, 5.2) -- (-0.5, 5.2);
\node[text width=3.1cm] at (1.1, 5.2) {\scriptsize \textit{Adv. tree of $P_{0}$}};	
\path [line, thick] (-1, 4.8) -- (-0.5, 4.8);
\node[text width=3.1cm] at (1.1, 4.8) {\scriptsize \textit{Adv. tree of $P_{1}$}};
\path [line, dotted, thick] (-1, 4.4) -- (-0.5, 4.4);
\node[text width=3.1cm] at (1.1, 4.4) {\scriptsize \textit{Adv. tree of $P_{2}$}};
\node[draw, circle, scale=\scaleSCOT, pattern=north west lines, pattern color=gray!40] (1) at (\x,\y) 			{$a,0$};
\node[draw, circle, scale=\scaleSCOT, pattern=north west lines, pattern color=gray!40] (2) at (\x + \xInc*1,\y) {$b,0$};
\node[draw, circle, scale=\scaleSCOT, pattern=north west lines, pattern color=gray!40] (3) at (\x+\xInc*2,\y)	{$c,0$};
\node[draw, circle, scale=\scaleSCOT, pattern=north west lines, pattern color=gray!40] (4) at (\x+\xInc*3,\y) 	{$d,0$};
\node[draw, circle, scale=\scaleSCOT, pattern=north west lines, pattern color=gray!40] (5) at (\x+\xInc*4,\y) 	{$e,0$};
\node[draw, circle, scale=\scaleSCOT, pattern=north west lines, pattern color=gray!40] (6) at (\x+\xInc*5+0.2,\y) 	{$f,0$};
\node[draw, circle, scale=\scaleSCOT, fill=green!50] (11) at (\x,\y-\yInc) 			{$a,1$};
\node[draw, circle, scale=\scaleSCOT, fill=green!50] (12) at (\x+ \xInc*1,\y-\yInc) {$b,1$};
\node[draw, circle, scale=\scaleSCOT, fill=green!50] (13) at (\x+ \xInc*2,\y-\yInc) {$c,1$};
\node[draw, circle, scale=\scaleSCOT, fill=green!50] (14) at (\x+ \xInc*3,\y-\yInc) {$d,1$};
\node[draw, circle, scale=\scaleSCOT, fill=green!50] (15) at (\x+ \xInc*4,\y-\yInc) {$e,1$};
\node[draw, circle, scale=\scaleSCOT, fill=green!50] (16) at (\x+ \xInc*5+0.2,\y-\yInc) {$f,1$};
\node[draw, circle, scale=\scaleSCOT] (21) at (\x,\y-\yInc*2) 		   {$a,2$};
\node[draw, circle, scale=\scaleSCOT] (22) at (\x+ \xInc*1,\y-\yInc*2) {$b,2$};
\node[draw, circle, scale=\scaleSCOT] (23) at (\x+ \xInc*2,\y-\yInc*2) {$c,2$};
\node[draw, circle, scale=\scaleSCOT] (24) at (\x+ \xInc*3,\y-\yInc*2) {$d,2$};
\node[draw, circle, scale=\scaleSCOT] (25) at (\x+ \xInc*4,\y-\yInc*2) {$e,2$};
\node[draw, circle, scale=\scaleSCOT] (26) at (\x+ \xInc*5+0.2,\y-\yInc*2) {$f,2$};
\node[draw, rectangle, scale=\scaleBox] (P0)	at  (\x-1,\y) {$P_{0}$};
\node[draw, rectangle, scale=\scaleBox] (P1)	at (\x+ \xInc*4+0.67, \y-\yInc*2+0.4) {$P_{1}$};
\node[draw, rectangle, scale=\scaleBox] (P2) at  (\x-1,\y-\yInc*2) {$P_{2}$};
\draw [thick, gray!70] (22) to [out=25,in=155] (24) (12) to [out=25,in=155] (14) (2) to [out=25,in=155] (4);
\draw [thick, gray!70] (1) -- (2) (2) -- (3) (4) -- (5) (5) -- (6) (11) -- (12) (12) -- (13) (14) -- (15) (15) -- (16) (21) -- (22) (22) -- (23) (24) -- (25) (25) -- (26);
\draw  (P2) -- (21) (P1) -- (15) (P0) -- (1); 
\draw  [dashed, thick, gray!70] (1) -- (11) (11) -- (21) (2) -- (12) (12) -- (22) (3) -- (13) (13) -- (23) (4) -- (14) (14) -- (24) 
(5) -- (15) (15) -- (25) (6) -- (16) (16) -- (26);
\draw [dashed, thick, gray!70] (1) to [out=240,in=120] (21) (2) to [out=240,in=120] (22) (3) to [out=240,in=120] (23) (4) to [out=240,in=120] 
(24) (5) to [out=240,in=120] (25) (6) to [out=240,in=120] (26);
\draw [line, thick, dashed] (P0) to [out=50,in=140] (1);
\draw [line, thick, dashed] (1) to [out=240,in=110] (11);
\draw [line, thick, dashed] (1) to [out=230,in=130] (21);
\draw [line, thick] (P1) to [out=90,in=340] (15);
\draw [line, thick] (15) to [out=50,in=310] (5); 
\draw [line, thick] (15) to [out=310,in=60] (25);
\draw [line, thick, dotted] (P2) to [out=30,in=150] (21);
\draw [line, thick, dotted] (21) to [out=60,in=300] (11);
\draw [line, thick, dotted] (21) to [out=60,in=300] (1);

\draw[color=blue, dotted] (\x-0.6,3.5) -- (\x-0.6,6.5);
\draw[color=blue, dotted] (\x+0.45,3.5) -- (\x+0.45,6.5);
\draw[color=blue, dotted] (\x+1.35,3.5) -- (\x+1.35,6.5);
\draw[color=blue, dotted] (\x+2.25,3.5) -- (\x+2.25,6.5);	
\draw[color=blue, dotted] (\x+3.15,3.5) -- (\x+3.15,6.5);	
\draw[color=blue, dotted] (\x+4.05,3.5) -- (\x+4.05,6.5);
\draw[color=blue, dotted] (\x+5,3.5) -- (\x+5,6.5);
\node[text width=2cm] at (\x+0.9,3.4) {\small $R_a$};
\node[text width=2cm] at (\x+1.8,3.4) {\small $R_b$};
\node[text width=2cm] at (\x+2.7,3.4) {\small $R_c$};
\node[text width=2cm] at (\x+3.6,3.4) {\small $R_d$};
\node[text width=2cm] at (\x+4.5,3.4) {\small $R_e$};
\node[text width=2cm] at (\x+5.6,3.4) {\small $R_f$};

\end{tikzpicture}

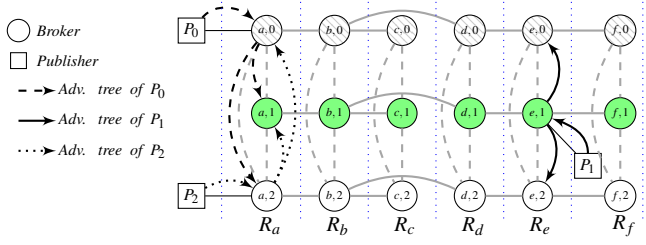
\captionof{figure}{Advertisements forwarding by three publisher $P_{0}$, $P_{1}$, and $P_{2}$. The host regions of $P_{0}$, and $P_{2}$ is $R_{a}$, and for $P_{1}$ is $R_{e}$.}
\section {Subscription Broadcast}
	\begin{tikzpicture}
	\def\scaleSCOT {0.5}
	\def\scaleBox {0.7}
	\tikzstyle{line} = [draw, -latex']
	\tikzstyle{lineR} = [draw, latex-']
	\node[draw, circle, scale=1] (lbroker) at (5.5,6.5) 			  {};
	\node[text width=1.5cm] at (6.5, 6.5) {\scriptsize \textit{Broker}};
	\node[draw, rectangle, scale=1] (lpub)	at  (5.5, 6.1) 		{};
	\node[text width=1.5cm] at (6.5, 6.1) {\scriptsize \textit{Subscriber}};		
	\path [line, dotted, thick] (5, 5.8) -- (5.5, 5.8);
	\node[text width=4cm] at (7.6, 5.8) {\scriptsize \textit{Sub. tree of $S$}};	
	\path [line, dashed, thick] (5, 5.4) -- (5.5, 5.4);
	\node[text width=4cm] at (7.6, 5.4) {\scriptsize \textit{Sub. tree of $S_{0}$}};	
	\path [line, blue, thick] (5, 5) -- (5.5, 5);
	\node[text width=4cm] at (7.6, 5) {\scriptsize \textit{Sub. tree of $S_{1}$}};	
	\path [line, thick] (5, 4.6) -- (5.5, 4.6);
	\node[text width=4cm] at (7.6, 4.6) {\scriptsize \textit{Sub. tree of $S_{2}$}};	
	
	\def\y {6}
	\def\x {-1}
	\def\xInc {1}
	\def\yInc {1.1}		
	\node[draw, circle, scale=\scaleSCOT, pattern=north west lines, pattern color=gray!40] (1) at (\x,\y) 				{$a,0$};
	\node[draw, circle, scale=\scaleSCOT, pattern=north west lines, pattern color=gray!40] (2) at (\x + \xInc*1,\y) {$b,0$};
	\node[draw, circle, scale=\scaleSCOT, pattern=north west lines, pattern color=gray!40] (3) at (\x+\xInc*2,\y)	{$c,0$};
	\node[draw, circle, scale=\scaleSCOT, pattern=north west lines, pattern color=gray!40] (4) at (\x+\xInc*3,\y) 	{$d,0$};
	\node[draw, circle, scale=\scaleSCOT, pattern=north west lines, pattern color=gray!40] (5) at (\x+\xInc*4,\y) 	{$e,0$};
	\node[draw, circle, scale=\scaleSCOT, pattern=north west lines, pattern color=gray!40] (6) at (\x+\xInc*5,\y) 	{$f,0$};
	\node[draw, circle, scale=\scaleSCOT, fill=green!50] (11) at (\x,\y-\yInc) 			{$a,1$};
	\node[draw, circle, scale=\scaleSCOT, fill=green!50] (12) at (\x+ \xInc*1,\y-\yInc) {$b,1$};
	\node[draw, circle, scale=\scaleSCOT, fill=green!50] (13) at (\x+ \xInc*2,\y-\yInc) {$c,1$};
	\node[draw, circle, scale=\scaleSCOT, fill=green!50] (14) at (\x+ \xInc*3,\y-\yInc) {$d,1$};
	\node[draw, circle, scale=\scaleSCOT, fill=green!50] (15) at (\x+ \xInc*4,\y-\yInc) {$e,1$};
	\node[draw, circle, scale=\scaleSCOT, fill=green!50] (16) at (\x+ \xInc*5,\y-\yInc) {$f,1$};
	\node[draw, circle, scale=\scaleSCOT] (21) at (\x,\y-\yInc*2) 		   	   {$a,2$};
	\node[draw, circle, scale=\scaleSCOT] (22) at (\x+ \xInc*1,\y-\yInc*2) {$b,2$};
	\node[draw, circle, scale=\scaleSCOT] (23) at (\x+ \xInc*2,\y-\yInc*2) {$c,2$};
	\node[draw, circle, scale=\scaleSCOT] (24) at (\x+ \xInc*3,\y-\yInc*2) {$d,2$};
	\node[draw, circle, scale=\scaleSCOT] (25) at (\x+ \xInc*4,\y-\yInc*2) {$e,2$};
	\node[draw, circle, scale=\scaleSCOT] (26) at (\x+ \xInc*5,\y-\yInc*2) {$f,2$};
	\node[draw, rectangle, scale=\scaleBox] (S0)	at  (\x,\y+0.8) 											 	{$S_{0}$};
	\node[draw, rectangle, scale=\scaleBox] (S)	at (\x+ \xInc*5, \y+0.8) 								 		{$S$};
	\node[draw, rectangle, scale=\scaleBox] (S1)	at (\x+ \xInc*3+0.6, \y-\yInc+0.7) 						{$S_{1}$};
	\node[draw, rectangle, scale=\scaleBox] (S2) at  (\x+ \xInc*5,\y-\yInc*2.5-0.2) 						 {$S_{2}$};
	\draw [thick, gray!70] (22) to [out=25,in=155] (24) (12) to [out=25,in=155] (14) (2) to [out=25,in=155] (4);
	\draw [thick, gray!70] (1) -- (2) (2) -- (3) (4) -- (5) (5) -- (6) (11) -- (12) (12) -- (13) (14) -- (15) (15) -- (16) (21) -- (22) (22) -- (23) (24) -- (25) (25) -- (26);
	\draw  [dashed, thick, gray!70] (1) -- (11) (11) -- (21) (2) -- (12) (12) -- (22) (3) -- (13) (13) -- (23) (4) -- (14) (14) -- (24) 
	(5) -- (15) (15) -- (25) (6) -- (16) (16) -- (26);
	\draw [dashed, thick, gray!70] (1) to [out=240,in=120] (21) (2) to [out=240,in=120] (22) (3) to [out=240,in=120] (23) (4) to [out=240,in=120] 
	(24) (5) to [out=240,in=120] (25) (6) to [out=240,in=120] (26);
	\draw [line, thick, dashed] (S0) to [out=330,in=45] (1);
	\draw [line, thick, dashed] (1) to [out=15,in=165] (2);
	\draw [line, thick, dashed] (2) to [out=15,in=165] (3);
	\draw [line, thick, dashed] (5) to [out=15,in=165] (6);
	\draw [line, thick, dashed] (4) to [out=15,in=165] (5);
	\draw [line, thick, dashed] (2) to [out=30,in=150] (4);
	\draw [line, thick, dotted] (S) to [out=210,in=120] (6);
	\draw [line, thick, dotted] (2) to [out=210,in=330] (1); 
	\draw [line, thick, dotted] (6) to [out=210,in=340] (5);
	\draw [line, thick, dotted] (5) to [out=210,in=340] (4); 
	\draw [line, thick, dotted] (4) to [out=210,in=330] (2);
	\draw [line, thick, dotted] (2) to [out=330,in=210] (3);
	\draw [line, thick, blue] (S1) to [out=180,in=80] (14);
	\draw [line, thick, blue] (12) to [out=210,in=330] (11); 
	\draw [line, thick, blue] (15) to [out=330,in=210] (16);
	\draw [line, thick, blue] (14) to [out=330,in=210] (15); 
	\draw [line, thick, blue] (14) to [out=210,in=330] (12);
	\draw [line, thick, blue] (12) to [out=330,in=210] (13);
	\draw [line, thick] (S2) to [out=30,in=320] (26);
	\draw [line, thick] (22) to [out=150,in=30] (21);
	\draw [line, thick] (22) to [out=30,in=150] (23);
	\draw [line, thick] (26) to [out=150,in=30] (25);
	\draw [line, thick] (25) to [out=150,in=30] (24);
	\draw [line, thick] (24) to [out=150,in=30] (22);
	
	\draw  (S0) -- (1) (S) -- (6); 
	\draw  (S1) -- (14); 
	\draw (S2) -- (26); 
	
	\draw [blue,dotted] (-1.5,6.4) -- (4.5,6.4);
	\draw [blue,dotted] (-1.5,5.4) -- (4.5,5.4);
	\draw [blue,dotted] (-1.5,4.3) -- (4.5,4.3);
	\draw [blue,dotted] (-1.5,3.5) -- (4.5,3.5);
	\node[text width=3cm] at (\x+6.8,6) {\small $C_0$};
	\node[text width=3cm] at (\x+6.8,4.9) {\small $C_1$};
	\node[text width=3cm] at (\x+6.8,3.8) {\small $C_2$};
	
	\end{tikzpicture}
	\captionof{figure}{Cluster--level subscription broadcast of subscribers $S$, $S_{0}$, $S_{1}$, and $S_{2}$. The host cluster of S and $S_{0}$ is $C_{0}$, while for $S_{1}$, and $S_{2}$ are $C_{1}$, and $C_{2}$ respectively.\\}
	To realize subgrouping and eliminate tight coupling, the \textit{Subscription Broadcast Process (SBP)} has to make sure that the subscriptions in a subgroup are confined to a single cluster. Therefore, SBP forwards a subscription to brokers in the host cluster of the subscriber, which is similar to subscription broadcast in traditional acyclic overlays \cite{SIENA_WIDE_AREA}. As shown in Fig. 5, subscription of $S_{0}$ is confined to brokers of its host cluster $C_{0}$ to form the subscription--tree. Similarly, subscription--trees of $S_{1}$ and $S_{2}$ are formed in their host clusters. SBP does not use trees of matching advertisements to generate publications routing paths \cite{SIENA_WIDE_AREA}. Instead, subscription-- and advertisement--trees are used independently for intra-- and inter--cluster routing, respectively. When a subscriber calls \textit{unsubscribe}, the corresponding subscription is simply removed from PRTs of brokers in the host cluster. This does not generate traffic in secondary clusters, which is even more effective when churn rate of subscribers is high.
\section{Cluster Index Vector}
A publisher may have interested subscribers in multiple clusters. As subscriptions in a subgroup are broadcast to only one cluster (and not available to other clusters), knowing the presence of interested subscribers in multiple clusters to send them publications is an issue. A naive solution suggests that publications should be forwarded on each iLink of the host broker of a publisher. However, this generates a number of false positives. 
\begin{tikzpicture}
\def\y {0.25}
\foreach \x in {2, 2.5, 3, 3.5}
\node[draw, rectangle, scale=2.1, thick] (B1)	at  (\x, 1) {};
\node at  (2, 1.5) {3};
\node at  (2.5, 1.5) {2};
\node at  (3, 1.5) {1};
\node at  (3.5, 1.5) {0};
\foreach \x in {2}
\node at  (\x, 1) {0};

\node[fill=black, text=white] at  (3, 1) {1};
\node[fill=gray!50] at  (2.5, 1) {1};
\node[fill=gray!50] at  (3.5, 1) {1};

\def\a {4.3}
\foreach \x in {0, 0.5, 1, 1.5}
\node[draw, rectangle, scale=2.1, thick] (B2)	at  (\x+\a, 1) {};
\node at  (0+\a, 1.5) {3};
\node at  (0.5+\a, 1.5) {2};
\node at  (1+\a, 1.5) {1};
\node at  (1.5+\a, 1.5) {0};

\foreach \x in {0, 1.5}
\node at  (\x+\a, 1) {0};

\node [fill=gray!50] at  (0.5+\a, 1) {1};
\node [fill=black, text=white] at (1+\a, 1) {1};
\node at  (1.5+\a, 1) {0};

\draw [decorate,decoration={brace,amplitude=4pt}, rotate=0, thick] (0 + \a, 1.7) -- (1.5 + \a, 1.7) node [black,midway,yshift=0.4cm] {\footnotesize \textit{Indexes}};
\draw [decoration={brace,mirror,raise=5pt, amplitude=4pt},decorate, thick] (-0.2+\a, 0.85) --  node[below=10pt]{\footnotesize \textit{Bit Values}}(1.7+\a, 0.85); 

\def\b {6.5}
\draw [decorate,decoration={brace,amplitude=4pt}, rotate=0, thick] (2, 1.7) -- (3.5, 1.7) node [black,midway,yshift=0.4cm] {\footnotesize \textit{Indexes}};
\draw [decoration={brace,mirror,raise=5pt, amplitude=4pt},decorate, thick] (1.8, 0.85) --  node[below=10pt]{\footnotesize \textit{Bit Values}} (3.7, 0.85); 
\foreach \x in {0, 0.5, 1, 1.5}
\node[draw, rectangle, scale=2.1, thick] (B2)	at  (\x+\b, 1) {};
\node at  (0+\b, 1.5) {3};
\node at  (0.5+\b, 1.5) {2};
\node at  (1+\b, 1.5) {1};
\node at  (1.5+\b, 1.5) {0};

\foreach \x in {0.5, 1.5}
\node[fill=gray!50] at  (\x+\b, 1) {1};

\node[fill=black, text=white] at  (1+\b, 1) {0};
\node at  (0+\b, 1) {0};

\draw [decorate,decoration={brace,amplitude=4pt}, rotate=0, thick] (0 + \b, 1.7) -- (1.5 + \b, 1.7) node [black,midway,yshift=0.4cm] {\footnotesize \textit{Indexes}};
\draw [decoration={brace,mirror,raise=5pt, amplitude=4pt},decorate, thick] (-0.2+\b, 0.85) --  node[below=10pt]{\footnotesize \textit{Bit Values}}(1.7+\b, 0.85); 

\node [fill=black, text=black] at (-0.2, 1.5) {\tiny 1};
\node[text width=2cm] at (1, 1.5) {\tiny \textit{Advertisement bit}};
\node [fill=gray!50, text=gray!50] at (-0.2, 1) {\tiny 1};
\node[text width=2cm] at (1, 1) {\tiny \textit{Subscription bit}};				

\node at (2.7, -0.2)  {\small (a) $CIV^1_{a}$};
\node at (5, -0.2)  {\small (b) $CIV^2_{s}$};
\node at (7.3, -0.2)  {\small (c) $CIV^1_{p}$};
\end{tikzpicture}
\captionof {figure}{(a) $CIV^1_{a}$ or $CIV^1_{a}[1]$ indicates that the publisher is hosted by a broker of $C_{1}$, while $C_{0}$ and $C_{2}$ are TSCs. (b) $CIV^2_{s}$ or $CIV^2_{s}[2]$ shows that interested subscribers are hosted by $C_{2}$. (c) $CIV^1_{p}[1]$ of the host cluster of the publisher is always 0 and $C_{2}$ and $C_{0}$ are TSCs.\\}
\textit{OctopiA} uses a novel concept of \textit{Cluster Index Vector (CIV)} to identify TSCs without forwarding subscriptions to other (non-host) clusters. CIV is a row (vector) of bits which identifies the host broker of a client, and the TSCs of a publisher. It is used to avoid false positives in static and dynamic routing of publications. Bits in \textit{CIV} are indexed from right to left, with the index of the right most bit being zero (Fig. 6). Each bit of $CIV$, called the (cluster) \textit{index bit}, is reserved for a cluster where index of the bit is same as index of the cluster it represents. Index bit of a broker is same as the index bit of the cluster it belongs to, while index of a client is the index of its host broker (or host cluster). Based on semantics, $CIV$ has three contexts: (i) \textit{advertisement context ($CIV^i_{a}$)}, (ii) \textit{subscription context ($CIV^j_{s}$)}, and (iii) \textit{publication context ($CIV^i_{p}$)}. $i \in V_{cf}$ is the index of the host cluster, while $j \in V_{cf}$ is the index of the secondary cluster of a publisher. $CIV^i_{a}$ and $CIV^j_{s}$ are associated with advertisements. More specifically, $CIV^i_{a}$ is the context of the advertisement at the host broker of the publisher, and changes to $CIV^j_{s}$ when the advertisement reaches secondary brokers. $CIV^i_{p}$ is used in dynamic routing of publications. Although maximum number of meaningful bits in $CIV$ are equal to the number of SCOT clusters, only two index bits in $CIV^j_{s}$ are meaningful. (i) $CIV^j_{s}[i]$ identifies index of the advertisement sending cluster $C_{i}$, and (ii) $CIV^j_{s}[j]$ is used by the advertisement receiving secondary cluster $C_{j}$. In Fig. 6(a),  $CIV^1_{a}[1]$ equal to 1 indicates that the publisher is connected with some broker of cluster $C_{1}$. In Fig. 1(b), $CIV^2_{s}[1]$ equal to 1 indicates that $C_{1}$ is the host cluster of the advertisement issuing publisher, while $CIV^2_{s}[2]$ shows that $C_{2}$ is the secondary cluster that received the advertisement. $CIV^1_{p}$ in Fig. 1(c) illustrates that the publisher is hosted by a broker of $C_{1}$, where $CIV^1_{p}[1]$ (or $CIV^i_{p}[i]$) has to be zero to avoid message loops in dynamic routing (Sec. 9). In the following, we discuss how index bits in CIV are set and reset to avoid false positives.\\
\textit{\textbf{CIV for Advertisement ---}}
After receiving an advertisement, the host broker of a publisher in its host cluster $C_{i}$ performs fours tasks: (i) creates $CIV^i_{a}$, (ii) sets $CIV^i_{a}[i]$ to 1 and all other index bits to 0, (iii) saves the advertisement in \textit{Cluster Link Table (CLT)} in the form of \textit{\{advertisement, $CIV^i_{a}$, lasthop\}} tuple, and (iv) forwards the advertisement with $CIV^i_{a}$ to secondary brokers. After receiving the advertisement, a secondary broker in a secondary cluster $C_{j}$ performs two tasks: (i) sets $CIV^j_{s}[j]$ to 1 if one or more matching subscriptions are found at cluster $C_{j}$, (ii) saves the advertisement in CLT in form of \textit{\{advertisement, $CIV^j_{s}$, lasthop\}} tuple. When $CIV^j_{s}[j]$ is set to 1, a $CIV\_SET$ message is sent to the host broker of the publisher in $C_{i}$ to indicate interested subscribers in $C_{j}$. The host broker sets the $CIV^i_{a}[j]$ to 1 to mark that $C_{j}$ is a TSC of the publisher. In Fig. 8, when \textit{B(e, 1)} receives an advertisement from $P_{1}$, it creates $CIV^1_{a}$, sets $CIV^1_{a}[1]$ to 1 and all other bits to 0, and forwards the advertisement (with $CIV^1_{a}$) to secondary brokers \textit{B(e, 0) and B(e,2)}. Assuming no matching subscriptions are found at clusters $C_{0}$ and $C_{2}$, both $CIV^0_{s}[0]$, and $CIV^2_{s}[2]$ are set to 0. In this scenario, $CIV^1_{a}$,  $CIV^0_{s}$, and $CIV^2_{s}$ are \textit{010}, which indicates that $P_{1}$ has no TSC. For $P_{2}$, we assume that cluster $C_{0}$ hosts interested subscribers. Secondary broker \textit{B(a, 0)} sets the $CIV^0_{s}[0]$ to 1 and sends a $CIV\_SET$ message to \textit{B(a, 2)} to indicate that $C_{0}$ is a TSC of $P_{2}$. Upon receiving the $CIV\_SET$ message, \textit{B(a, 2)} sets $CIV^2_{a}[0]$ to 1. $CIV^2_{a}$ for the advertisement issued by $P_{2}$ is $101$.

Algorithm 1 shows that the host broker of the publisher sets the bits and forwards $CIV^i_{a}$ and the advertisement to secondary brokers (lines 2--8). Each secondary broker sends only one $CIV\_SET$ message even if multiple matching subscriptions are found (lines 14--20). The host broker of the publisher sets the $CIV^i_{a}[j]$ to 1 when a $CIV\_SET$ message is received from a broker in $C_{j}$ cluster (line 11). Each advertisement has a unique identification \textit{(UID)}, which is used to identify the advertisement whose $CIV^i_{a}$ should be updated (line 10). 
\begin {algorithm}
\footnotesize
\KwIn{$a:$ an advertisement message\;}
\KwOut{$DL:$ a list of next destinations that should receive $a$\;}
\BlankLine
\textit{/* set the civ bits at the host broker of the publisher */}\;
\If{$isHostBroker (a)$}{
	$a.CIV^i_{a}[i] \gets 1$\;
	$a.hostBroker \gets false$\;
	\ForEach {$n \in SecondaryNeighbours$}{
		$a.next \gets n$\;
		$DL.add(a)$\;		
	}
	$CLT.insert(a)$\;
} \ElseIf {$a.type = CIV\_SET$}{
$savedAdv \gets getAdv (a.UID)$\;		
$savedAdv.CIV^i_{a}[j] \gets a.CIV^j_{s}[j]$\;
}\Else{
\BlankLine
\textit{/* send CIV\_SET message if matching subscription is found */}\;
\If{$isMatchingSub(a) = true$}{
	$a.next \gets a.last$\;
	$a.type \gets CIV\_SET$\;
	$a.CIV^j_{s}[j] \gets 1$\;
	$DL.add(a)$\;	
}
$CLT.insert(a)$\;
}
\BlankLine		
\label{algo:AFP}
\caption{$scotAFP(a)$}
\end{algorithm}
\textit{\textbf{CIV for Subscription ---}}
Upon receiving a new subscription, a secondary broker $B(a, j)$ of a secondary cluster $C_{j}$ sends $CIV\_SET$ message to the host broker $B(b, i)$ of the publisher in it host cluster $C_{i}$ \textit{iff} the corresponding advertisement matches with the subscription and $CIV^j_{s}[j]$ is 0. After sending $CIV\_SET$ message, $B(a, j)$ sets $CIV^j_{s}[j]$ to 1 to stop sending $CIV\_SET$ messages when more new subscriptions matching with the advertisement are received. After receiving $CIV\_SET$ message, $B(b, i)$ sets $CIV^i_{a}[j]$ to 1 to indicate that $C_{j}$ is a TSC of the publisher. $CIV\_SET$ message also carries the new subscription to avoid sending multiple $CIV\_SET$ messages when multiple matching advertisements are received from the same host broker. Subscriptions received with a $CIV\_SET$ messages are not saved in routing tables but used to find matching advertisements for which the index bits $CIV^i_{a}[j]$ are 0. In Fig. 8, assuming that $S_{2}$ issues the first subscription that matches with the advertisement from $P_{0}$, \textit{B(a, 2)} sets $CIV^2_{s}[2]$ to 1 and sends a $CIV\_SET$ message (with the subscription from $S_{2}$) to the host broker \textit{B(a, 0)}, which sets $CIV^0_{a}[2]$ to 1 to indicate that cluster $C_{2}$ is a TSC of $P_{0}$. Fig. 6(a) indicates that clusters $C_{0}$, and $C_{2}$ are TSCs of the publisher hosted by a broker of cluster $C_{1}$. Similarly, Fig. 6(b) shows that the advertisement is received from a broker of cluster $C_{2}$ by a publisher hosted by a broker of cluster $C_{1}$, while $C_{2}$ host one or more interested subscribers as $CIV^2_{s}$ is 1. A $CIV\_UNSET$ message is sent to the host broker of a publisher when the last interested subscriber in a TSC $C_{j}$ calls \textit{unsubscribe}. After receiving the $CIV\_UNSET$ message, the index bit $CIV^i_{a}[j]$ is set to 0 to avoid false positives.

Algorithm 2 indicates that a subscription is broadcast to brokers in its host cluster (lines 1-3). $getDistinctMatchingAdvs$ returns at most one matching advertisement per iLink for which $CIV^j_{s}[j]$ is 0. $CIV\_SET$ message is sent to each secondary broker to set $CIV^i_{a}[j]$ to 1 (lines 6-10). 

\begin {algorithm}
\footnotesize
\KwIn{$s:$ a subscription message\;}
\KwOut{$DL:$ a list of next destinations that should receive $s$\;}

\ForEach {$n \in (PrimaryNeighbours - Sender)$}{
	$s.next \gets n$\;
	$DL.add(s)$\;
}

$PRT.insert(s)$\;	

\textit{/* select distinct advertisements which have $CIV^j_{s}[j]$ 0. */}\;
$advs \gets getDistinctMatchingAdvs(s)$\;

\ForEach {$a \in advs$}{
	$CIV\_SET.next \gets a.last$\;
	$CIV\_SET.sub \gets s$\;
	$DL.add(CIV\_SET)$\;
}
\label{algo:A1}
\caption{$scotSBP(s)$}
\end{algorithm}
\section{Publications Routing}
\begin{figure*}
	\tikzstyle{line} = [draw, -latex']
	\tikzstyle{line} = [draw, -latex']
	\centering
	\def\scaleSCOT {0.5}
	\def\scaleBox {0.6}
	\def\x {-1} 
	\def\xInc {1.1}
	\def\y {0} 
	\def\yInc {0.9}
	\def\gr {50}
	
	\begin{subfigure}[b]{0.27\textwidth}
		\captionsetup{width=0.95\linewidth}
		\begin{tikzpicture}			
		\tikzstyle{every node} = [thick]
		\node[draw, circle, scale=\scaleSCOT+0.2] (1) at (\x-0.5,\y) {$a$};
		\node[draw, circle, scale=\scaleSCOT+0.15] (2) at (\x-0.65+\xInc,\y) {$b$};
		\node[draw, circle, scale=\scaleSCOT+0.25] (3) at (\x-0.8+\xInc*2,\y) {$c$};
		\node[draw, thick, rectangle, scale=1.1] (op)	at (\x-0.2+\xInc*1.9,\y) {};
		\draw [gray!70, thick] (1) -- (2) (2) -- (3);
		\node[draw, circle, scale=\scaleSCOT+0.18] (va) at (\x+\xInc*2.2,\y) {$1$};
		\node[draw, circle, scale=\scaleSCOT+0.18] (vb) at (\x+0.7+\xInc*3,\y) {$2$};
		\node[draw, circle, scale=\scaleSCOT+0.18] (vc) at (\x+\xInc*3-0.05,\y+1.3) {$0$};
		\node[draw, circle, scale=\scaleSCOT+0.18] (vd) at (\x+\xInc*3-0.05,\y+0.45) {$3$};		
		\draw [dashed, thick] (va) -- (vb) (vb) -- (vc) (vc) -- (va) (vd) -- (va) (vd) -- (vb) (vd) -- (vc);
		\path [line, thick] (-1.2, 3.2) -- (-0.7, 3.2);
		\node[text width=3.5cm] at (1.2, 3.2) {\tiny \textit{Intra-cluster publiation routing}};
		\path [line, thick, dashed] (-1.2, 2.9) -- (-0.7, 2.9);
		\node[text width=3.5cm] at (1.2, 2.9) {\tiny \textit{Inter-cluster publication routing}};
		\path [line, thick, red] (-1.2, 2.6) -- (-0.7, 2.6);
		\node[text width=4cm] at (1.4, 2.6) {\tiny \textit{Intra-cluster $CIV^i_{p}-D$ routing}};
		\path [line, thick, red, dashed] (-1.2, 2.3) -- (-0.7, 2.3);
		\node[text width=4cm] at (1.4, 2.3) {\tiny \textit{Inter-cluster $CIV^i_{p}-D$ routing}};
		\draw [line width=0.08cm, gray!50] (-1.2, 2) -- (-0.7, 2);
		\node[text width=3.5cm] at (1.2, 2) {\tiny \textit{Overloaded aLink}};
		\draw [line width=0.08cm, gray!50, dashed] (-1.2, 1.7) -- (-0.7, 1.7);
		\node[text width=3.5cm] at (1.2, 1.7) {\tiny \textit{Overloaded iLink}};
		\node[draw, circle, scale=1] (lbroker) at (-1,1) 			  {};
		\node[text width=1.5cm] at (0, 1) {\tiny \textit{Broker}};
		\node[draw, rectangle, scale=1] (lpub)	at  (-1, 1.4) 		{};
		\node[text width=3cm] at (0.7, 1.4) {\tiny \textit{Publisher or Subscriber}};		
		\end{tikzpicture}
		\caption{\footnotesize Operands of a SCOT: tetrahedron ($G_{cf}$) generates four replicas of a 3-nodes path graph ($G_{af}$). }
		\label{fig:IDR2}
	\end{subfigure}
	~
	\begin{subfigure}[b]{0.23\textwidth}
		\captionsetup{width=0.95\linewidth}
		\begin{tikzpicture}
		\tikzstyle{every node} = [thick]
		\node[draw, circle, scale=\scaleSCOT] (1) at (\x,\y+\yInc*2) 						{$a,0$};
		\node[draw, circle, scale=\scaleSCOT] (2) at (\x+\xInc,\y+\yInc*2) 					{$b,0$};
		\node[draw, circle, scale=\scaleSCOT] (3) at (\x+\xInc*2,\y+\yInc*2) 				{$c,0$};
		\node[draw, circle, scale=\scaleSCOT] (4) at (\x,\y+\yInc) 							{$a,1$};
		\node[draw, circle, scale=\scaleSCOT] (5) at (\x+\xInc,\y+\yInc) 					{$b,1$};
		\node[draw, circle, scale=\scaleSCOT] (6) at (\x+\xInc*2,\y+\yInc) 					{$c,1$};
		\node[draw, circle, scale=\scaleSCOT] (7) at (\x,\y) 								{$a,2$};
		\node[draw, circle, scale=\scaleSCOT] (8) at (\x+\xInc,\y) 							{$b,2$};
		\node[draw, circle, scale=\scaleSCOT] (9) at (\x+\xInc*2,\y) 						{$c,2$};
		\node[draw, circle, scale=\scaleSCOT] (10) at (\x,\y-\yInc) 						{$a,3$};
		\node[draw, circle, scale=\scaleSCOT] (11) at (\x+\xInc,\y-\yInc) 					{$b,3$};
		\node[draw, circle, scale=\scaleSCOT] (12) at (\x+\xInc*2,\y-\yInc) 				{$c,3$};
		\node[draw, rectangle, scale=\scaleBox] (S1)	at (\x+\xInc*2+1,\y+\yInc*2)  		{$S1$};
		\node[draw, rectangle, scale=\scaleBox] (P)		at (\x+\xInc,\y-\yInc*2+0.2) 			{$P$};
		\node[draw, rectangle, scale=\scaleBox] (S2)	at (\x+\xInc*2+1, \y+\yInc)			{$S2$};
		\node[draw, rectangle, scale=\scaleBox] (S3)	at (\x,\y-\yInc*2+0.2) 					{$S3$};
		\node[draw, rectangle, scale=\scaleBox] (S4)	at (\x+\xInc*2,\y-\yInc*2+0.2) 			{$S4$};
		\node[draw, rectangle, scale=\scaleBox] (S5)	at (\x+\xInc*2+1,\y)		 		{$S5$};			
		\draw [gray!\gr, thick] (1) -- (2) (2) -- (3) (4) -- (5) (5) -- (6) (7) -- (8) (8) -- (9) ;
		\draw [dashed, gray!\gr, thick] (1) -- (4) (4) -- (7) (3) -- (6) (6) -- (9);
		\draw [thick] (P) -- (11) (S1) -- (3) (S2) -- (6) (S3) -- (10) (S4) -- (12) (S5) -- (9);
		\draw [dashed, gray!\gr, thick] (7) -- (10) (8) -- (11) (9) -- (12) (5) -- (8);
		\draw [gray!\gr, thick] (10) -- (11) (11) -- (12);
		\draw [line width=0.08cm, dashed, gray!\gr] (2) -- (5);			
		\draw [dashed, gray!\gr, thick]  (1) to [out=230,in=130] (10);
		\draw [line width=0.08cm, dashed, gray!\gr]  (2) to [out=230,in=130] (11);
		\draw [dashed, gray!\gr, thick]  (3) to [out=230,in=130] (12);			
		\draw [dashed, thick, gray!\gr]  (1) to [out=240,in=120] (7);
		\draw [dashed, thick, gray!\gr]  (4) to [out=240,in=120] (10);			
		\draw [dashed, thick, gray!\gr]  (2) to [out=240,in=120] (8);
		\draw [dashed, gray!\gr, thick]  (5) to [out=240,in=120] (11); 			
		\draw [dashed, thick, gray!\gr]  (3) to [out=240,in=120] (9);
		\draw [dashed, thick, gray!\gr]  (6) to [out=240,in=120] (12);			
		\path [line, thick] (P) to [out=60,in=290] (11);
		\path [line, thick] (11) to [out=40,in=150] (12);
		\path [line, thick] (8) to [out=40,in=150] (9);
		\path [line, thick] (6) to [out=40,in=150] (S2);
		\path [line, thick] (9) to [out=40,in=150] (S5);
		\path [line, thick] (11) to [out=140,in=30] (10);
		\path [line, thick] (3) to [out=30,in=150] (S1);
		\path [line, thick] (12) to [out=250,in=110] (S4);
		\path [line, thick] (10) to [out=250,in=110] (S3);
		\path [line, dashed, thick] (11) to [out=70,in=290] (8);
		\path [line, dashed, thick, red] (11) to [out=60,in=300] (5);
		\path [line, thick, red] (5) to [out=40,in=150] (6);
		\path [line, dashed, thick] (6) to [out=70,in=290] (3);				 				
		\end{tikzpicture}
		\caption{\footnotesize Two iLinks, $l \langle(b,3), (b,0) \rangle$ and $l \langle(b, 1), (b, 0) \rangle$, are congested.}
		\label{fig:DNR2}
	\end{subfigure}
	~
	\begin{subfigure}[b]{0.23\textwidth}
		\captionsetup{width=0.95\linewidth}
		\begin{tikzpicture}
		\tikzstyle{every node} = [thick]
		\node[draw, circle, scale=\scaleSCOT] (1) at (\x,\y+\yInc*2) 						{$a,0$};
		\node[draw, circle, scale=\scaleSCOT] (2) at (\x+\xInc,\y+\yInc*2) 					{$b,0$};
		\node[draw, circle, scale=\scaleSCOT] (3) at (\x+\xInc*2,\y+\yInc*2) 				{$c,0$};
		\node[draw, circle, scale=\scaleSCOT] (4) at (\x,\y+\yInc) 							{$a,1$};
		\node[draw, circle, scale=\scaleSCOT] (5) at (\x+\xInc,\y+\yInc) 					{$b,1$};
		\node[draw, circle, scale=\scaleSCOT] (6) at (\x+\xInc*2,\y+\yInc) 					{$c,1$};
		\node[draw, circle, scale=\scaleSCOT] (7) at (\x,\y) 								{$a,2$};
		\node[draw, circle, scale=\scaleSCOT] (8) at (\x+\xInc,\y) 							{$b,2$};
		\node[draw, circle, scale=\scaleSCOT] (9) at (\x+\xInc*2,\y) 						{$c,2$};
		\node[draw, circle, scale=\scaleSCOT] (10) at (\x,\y-\yInc) 						{$a,3$};
		\node[draw, circle, scale=\scaleSCOT] (11) at (\x+\xInc,\y-\yInc) 					{$b,3$};
		\node[draw, circle, scale=\scaleSCOT] (12) at (\x+\xInc*2,\y-\yInc) 				{$c,3$};
		\node[draw, rectangle, scale=\scaleBox] (S1)	at (\x+\xInc*2+0.8,\y+\yInc*2)  		{$S1$};
		\node[draw, rectangle, scale=\scaleBox] (P)		at (\x+\xInc,\y-\yInc*2+0.2) 			{$P$};
		\node[draw, rectangle, scale=\scaleBox] (S2)	at (\x-\xInc+0.3,\y+\yInc*2)			{$S2$};
		\node[draw, rectangle, scale=\scaleBox] (S3)	at (\x,\y-\yInc*2+0.2) 					{$S3$};
		\node[draw, rectangle, scale=\scaleBox] (S4)	at (\x+\xInc*2,\y-\yInc*2+0.2) 			{$S4$};
		\node[draw, rectangle, scale=\scaleBox] (S5)	at (\x+\xInc*2+0.8,\y)		 		{$S5$};			
		\draw [gray!\gr, thick] (1) -- (2) (2) -- (3) (4) -- (5) (5) -- (6) (7) -- (8) (8) -- (9) ;
		\draw [dashed, gray!\gr, thick] (1) -- (4) (4) -- (7) (2) -- (5) (3) -- (6) (6) -- (9);
		\draw [thick] (P) -- (11) (S1) -- (3) (S2) -- (1) (S3) -- (10) (S4) -- (12) (S5) -- (9);
		\draw [dashed, gray!\gr, thick] (7) -- (10) (9) -- (12) (5) -- (8);
		\draw [gray!\gr, thick] (10) -- (11) (11) -- (12);
		\draw [line width=0.08cm, dashed, gray!\gr] (8) -- (11);
		\draw [dashed, gray!\gr, thick]  (1) to [out=230,in=130] (10);
		\draw [line width=0.08cm, dashed, gray!\gr]  (2) to [out=230,in=130] (11);
		\draw [thick, dashed, gray!\gr]  (3) to [out=230,in=130] (12);		
		\draw [dashed, thick, gray!\gr]  (1) to [out=240,in=120] (7);
		\draw [dashed, thick, gray!\gr]  (4) to [out=240,in=120] (10);		
		\draw [dashed, thick, gray!\gr]  (2) to [out=240,in=120] (8);
		\draw [dashed, thick, gray!\gr]  (5) to [out=240,in=120] (11); 		
		\draw [dashed, thick, gray!\gr]  (3) to [out=240,in=120] (9);
		\draw [dashed, thick, gray!\gr]  (6) to [out=240,in=120] (12);		
		\path [line, thick] (P) to [out=60,in=290] (11);
		\path [line, thick, red] (11) to [out=30,in=150] (12);
		\path [line, thick] (9) to [out=30,in=150] (S5);
		\path [line, thick] (11) to [out=150,in=30] (10);
		\path [line, thick] (3) to [out=150,in=30] (2);
		\path [line, thick] (1) to [out=150,in=30] (S2);		
		\path [line, thick] (2) to [out=150,in=30] (1);		
		\path [line, thick] (3) to [out=30,in=150] (S1);
		\path [line, thick] (12) to [out=250,in=110] (S4);
		\path [line, thick] (10) to [out=250,in=110] (S3);		
		\path [line, dashed, thick] (12) to [out=70,in=290] (9);
		\path [line, dashed, thick] (12) to [out=60,in=300] (3);				 				
		\end{tikzpicture}
		\caption{\footnotesize Two iLinks, $l \langle(b,3), (b,0) \rangle$ and $l \langle(b, 3), (b, 2) \rangle$, are congested.}
		\label{fig:DNR1}
	\end{subfigure}
	~
	\begin{subfigure}[b]{0.23\textwidth}
		\captionsetup{width=0.95\linewidth}		
		\begin{tikzpicture}
		\tikzstyle{every node} = [thick]
		\node[draw, circle, scale=\scaleSCOT] (1) at (\x,\y+\yInc*2) 						{$a,0$};
		\node[draw, circle, scale=\scaleSCOT] (2) at (\x+\xInc,\y+\yInc*2) 					{$b,0$};
		\node[draw, circle, scale=\scaleSCOT] (3) at (\x+\xInc*2,\y+\yInc*2) 				{$c,0$};
		\node[draw, circle, scale=\scaleSCOT] (4) at (\x,\y+\yInc) 							{$a,1$};
		\node[draw, circle, scale=\scaleSCOT] (5) at (\x+\xInc,\y+\yInc) 					{$b,1$};
		\node[draw, circle, scale=\scaleSCOT] (6) at (\x+\xInc*2,\y+\yInc) 					{$c,1$};
		\node[draw, circle, scale=\scaleSCOT] (7) at (\x,\y) 								{$a,2$};
		\node[draw, circle, scale=\scaleSCOT] (8) at (\x+\xInc,\y) 							{$b,2$};
		\node[draw, circle, scale=\scaleSCOT] (9) at (\x+\xInc*2,\y) 						{$c,2$};
		\node[draw, circle, scale=\scaleSCOT] (10) at (\x,\y-\yInc) 						{$a,3$};
		\node[draw, circle, scale=\scaleSCOT] (11) at (\x+\xInc,\y-\yInc) 					{$b,3$};
		\node[draw, circle, scale=\scaleSCOT] (12) at (\x+\xInc*2,\y-\yInc) 				{$c,3$};
		\node[draw, rectangle, scale=\scaleBox] (S1)	at (\x-\xInc+0.4,\y+\yInc*2)  		{$S1$};
		\node[draw, rectangle, scale=\scaleBox] (P)		at (\x+\xInc,\y-\yInc*2+0.2) 			{$P$};
		\node[draw, rectangle, scale=\scaleBox] (S2)	at (\x+\xInc*2+1, \y+\yInc)			{$S2$};
		\node[draw, rectangle, scale=\scaleBox] (S3)	at (\x+\xInc*2+0.5,\y-\yInc*2+0.2) 	{$S3$};
		\node[draw, rectangle, scale=\scaleBox] (S4)	at (\x+\xInc*2,\y-\yInc*2+0.2) 			{$S4$};
		\node[draw, rectangle, scale=\scaleBox] (S5)	at (\x+\xInc*2+1,\y)		 		{$S5$};						
		\draw [gray!\gr, thick] (1) -- (2) (2) -- (3) (4) -- (5) (5) -- (6) (7) -- (8) (8) -- (9) ;
		\draw [dashed, gray!\gr, thick] (4) -- (7) (3) -- (6) (6) -- (9);
		\draw [thick] (P) -- (11) (S1) -- (1) (S2) -- (6) (S3) -- (12) (S4) -- (12) (S5) -- (9);
		\draw [dashed, gray!\gr, thick] (7) -- (10) (9) -- (12) (5) -- (8);
		\draw [gray!\gr, thick] (10) -- (11);			
		\draw [dashed, gray!\gr, thick]  (1) to [out=230,in=130] (10);
		\draw [dashed, gray!\gr, thick]  (3) to [out=230,in=130] (12);
		\draw [dashed, thick, gray!\gr]  (1) to [out=240,in=120] (7);
		\draw [dashed, thick, gray!\gr]  (4) to [out=240,in=120] (10);		
		\draw [dashed, thick, gray!\gr]  (2) to [out=240,in=120] (8);
		\draw [line width=0.08cm, dashed, gray!\gr]  (5) to [out=240,in=120] (11); 
		\draw [line width=0.08cm, dashed, gray!\gr]  (2) to [out=230,in=130] (11);
		\draw [line width=0.08cm, gray!\gr]  (12)-- (11) ;
		\draw [line width=0.08cm, gray!\gr, dashed] (11)--(8) (5) -- (8) (2)-- (5) (1) -- (4);		
		\draw [dashed, thick, gray!\gr]  (3) to [out=240,in=120] (9);
		\draw [dashed, thick, gray!\gr]  (6) to [out=240,in=120] (12);
		\path [line, thick] (P) to [out=60,in=290] (11);
		\path [line, thick] (11) to [out=30,in=150] (12);
		\path [line, thick] (6) to [out=30,in=150] (S2);
		\path [line, thick] (9) to [out=30,in=150] (S5);
		\path [line, thick] (1) to [out=150,in=30] (S1);
		\path [line, thick] (12) to [out=250,in=110] (S4);
		\path [line, thick] (12) to [out=330,in=100] (S3);
		\path [line, dashed, thick, red] (11) to [out=70,in=290] (8);
		\path [line, dashed, thick] (1) to [out=250,in=110] (4);
		\path [line, dashed, red, thick] (8) to [out=60,in=300] (2);
		\path [line, thick] (8) to [out=30,in=150] (9);
		\path [line, thick] (5) to [out=30,in=150] (6);
		\path [line, thick] (4) to [out=30,in=150] (5);
		\path [line, thick, red] (2) to [out=150,in=30] (1);												 				
		\end{tikzpicture}
		\caption{\footnotesize All target links (3 iLinks, 1 aLink) at \textit{B(b, 3)} are congested. This is a padding...}
		\label{fig:DNR3}	
	\end{subfigure}
	\caption{A SCOT overlay of 12 brokers (4 clusters, 3 regions). The set of subscribers \textit{\{S1, S2, S3, S4\}} is interested in publications from a publisher P hosted by the broker B(b, 3).} 
	\label{fig:DNR}
\end{figure*}
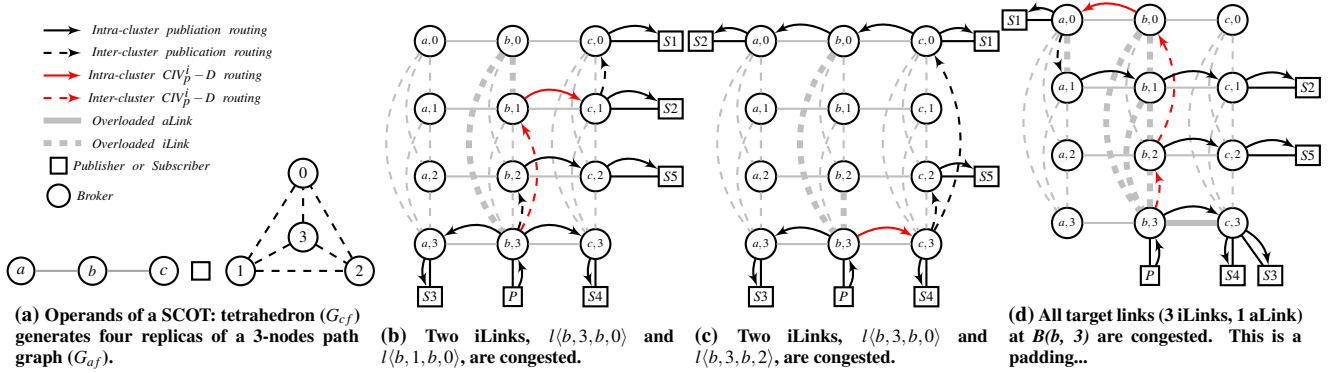
This section discusses algorithms for static and dynamic publication routing in \texttt{OctopiA}.
\subsection{Static Routing}
\textit{Static Publication Routing (SPR)} algorithm uses advertisement--trees of length 1 and cluster--level subscription--trees for publication routing in a clustered SCOT. The algorithm deals with two scenarios. (i) When the host cluster of a publisher is the only target cluster, and lengths of the routing paths satisfy the relation:
\begin{center}
	$max\big(d\langle(u_{1},v_{1}), (u_{2}, v_{2})\rangle \big) \leq diam(G_{af})$
\end{center}
where $(u_{1},v_{1})$ and $(u_{2}, v_{2})$ are any brokers in the same cluster. No inter--cluster messaging takes place as no publication is forwarded to secondary brokers because all $CIV^i_{a}[j]$, for secondary clusters, are 0. (ii) When at least one interested subscriber is hosted by a secondary cluster and therefore lengths of the routing paths satisfy the relation:
\begin{center}
	$max\big(d\langle(x_{1},y_{1}), (x_{2}, y_{2})\rangle \big) \leq \big( diam(G_{af}) + 1 \big)$
\end{center} 
where $(x_{1},y_{1})$ and $(x_{2}, y_{2})$ are any brokers in a clustered SCOT. The host broker of a publisher forwards publications onto target iLinks (recall that a target link is part of a routing path to next destination) for which index bits $CIV^i_{a}[j]$, where $j \in V_{cf}$, are 1. This avoids false positives by sending publications to those clusters which have interested subscribers. After the first broker of a TSC receives the publications, only intra--cluster routing is used to forward the publications to interested subscribers in a TSC. No loops appear and no path identifications are required because each cluster is an acyclic overlay. In Fig. 8, $CIV^i_{a}$ for advertisements from $P_{0}$, $P_{1}$, and $P_{2}$ are \textit{110, 111}, and \textit{100}, respectively. For $P_{0}$ and $P_{2}$, routing is only inter-- and intra--cluster, respectively. Publications from $P_{1}$ result in both intra-- and inter--cluster routing as the publisher has interested subscribers in all three clusters. \textit{B(e, 1)} forwards publications to primary broker \textit{B(d, 1)} therefore messaging onto the aLink $l \langle(e,1), (d,1) \rangle$ is intra--cluster, while publications routing to secondary brokers \textit{B(e,0)} and \textit{B(e, 2)} onto the iLinks $l \langle(e,1), (e,0) \rangle$ and $l \langle(e,1), (e,2) \rangle$ is inter--cluster.
\begin{tikzpicture}
\def\scaleSCOT {0.55}
\def\scaleBox {0.8}
\tikzstyle{line} = [draw, -latex']
\tikzstyle{lineR} = [draw, latex-']
\def\y {6}
\def\x {0}
\def\xInc {1.1}
\def\yInc {1.2}		
\node[draw, circle, scale=\scaleSCOT, pattern=north west lines, pattern color=gray!40] (1) at (\x,\y) 			{$a,0$};
\node[draw, circle, scale=\scaleSCOT, pattern=north west lines, pattern color=gray!40] (2) at (\x + \xInc*1,\y) {$b,0$};
\node[draw, circle, scale=\scaleSCOT, pattern=north west lines, pattern color=gray!40] (3) at (\x+\xInc*2,\y)	{$c,0$};
\node[draw, circle, scale=\scaleSCOT, pattern=north west lines, pattern color=gray!40] (4) at (\x+\xInc*3,\y) 	{$d,0$};
\node[draw, circle, scale=\scaleSCOT, pattern=north west lines, pattern color=gray!40] (5) at (\x+\xInc*4,\y) 	{$e,0$};
\node[draw, circle, scale=\scaleSCOT, pattern=north west lines, pattern color=gray!40] (6) at (\x+\xInc*5+0.5,\y) 	{$f,0$};
\node[draw, circle, scale=\scaleSCOT, fill=green!50] (11) at (\x,\y-\yInc) 			{$a,1$};
\node[draw, circle, scale=\scaleSCOT, fill=green!50] (12) at (\x+ \xInc*1,\y-\yInc) {$b,1$};
\node[draw, circle, scale=\scaleSCOT, fill=green!50] (13) at (\x+ \xInc*2,\y-\yInc) {$c,1$};
\node[draw, circle, scale=\scaleSCOT, fill=green!50] (14) at (\x+ \xInc*3,\y-\yInc) {$d,1$};
\node[draw, circle, scale=\scaleSCOT, fill=green!50] (15) at (\x+ \xInc*4,\y-\yInc) {$e,1$};
\node[draw, circle, scale=\scaleSCOT, fill=green!50] (16) at (\x+ \xInc*5+0.5,\y-\yInc) {$f,1$};
\node[draw, circle, scale=\scaleSCOT] (21) at (\x,\y-\yInc*2) 		   {$a,2$};
\node[draw, circle, scale=\scaleSCOT] (22) at (\x+ \xInc*1,\y-\yInc*2) {$b,2$};
\node[draw, circle, scale=\scaleSCOT] (23) at (\x+ \xInc*2,\y-\yInc*2) {$c,2$};
\node[draw, circle, scale=\scaleSCOT] (24) at (\x+ \xInc*3,\y-\yInc*2) {$d,2$};
\node[draw, circle, scale=\scaleSCOT] (25) at (\x+ \xInc*4,\y-\yInc*2) {$e,2$};
\node[draw, circle, scale=\scaleSCOT] (26) at (\x+ \xInc*5+0.5,\y-\yInc*2) {$f,2$};
\node[draw, rectangle, scale=\scaleBox] (S0)	at  (\x-1,\y) 					{$S_{0}$};
\node[draw, rectangle, scale=\scaleBox] (S)	at (\x+ \xInc*5+1.5, \y) 		{$S$};
\node[draw, rectangle, scale=\scaleBox] (S1)	at (\x+ \xInc*3+0.6, \y-\yInc+0.8) 	{$S_{1}$};
\node[draw, rectangle, scale=\scaleBox] (S2) at  (\x+ \xInc*5+1.5,\y-\yInc*2) 			{$S_{2}$};
\node[draw, rectangle, scale=\scaleBox] (P2)	at (\x-1, \y-\yInc*2) 			{$P_{2}$};
\node[draw, rectangle, scale=\scaleBox] (P0)	at (\x-1, \y-\yInc+0.5) 			{$P_{0}$};
\node[draw, rectangle, scale=\scaleBox] (P1) at  (\x+ \xInc*5-0.2,\y-\yInc*1.5)	 		{$P_{1}$};	
\draw [thick, gray!70] (22) to [out=25,in=155] (24) (12) to [out=25,in=155] (14) (2) to [out=25,in=155] (4);
\draw [thick, gray!70] (1) -- (2) (2) -- (3) (4) -- (5) (5) -- (6) (11) -- (12) (12) -- (13) (14) -- (15) (15) -- (16) (21) -- (22) (22) -- (23) (24) -- (25) (25) -- (26);
\draw  [dashed, thick, gray!70] (1) -- (11) (11) -- (21) (2) -- (12) (12) -- (22) (3) -- (13) (13) -- (23) (4) -- (14) (14) -- (24) 
(5) -- (15) (15) -- (25) (6) -- (16) (16) -- (26);
\draw [dashed, thick, gray!70] (1) to [out=240,in=120] (21) (2) to [out=240,in=120] (22) (3) to [out=240,in=120] (23) (4) to [out=240,in=120] 
(24) (5) to [out=240,in=120] (25) (6) to [out=240,in=120] (26);
\draw [line, thick, dotted] (P1) to [out=170,in=310] (15);
\draw [line, thick, dotted] (15) to [out=140,in=30] (14); 
\draw [line, thick, dotted] (14) to [out=90,in=190] (S1);
\draw [line, thick, dotted] (15) to [out=60,in=300] (5); 
\draw [line, thick, dotted] (15) to [out=290,in=70] (25);
\draw [line, thick, dotted] (25) to [out=30,in=150] (26);
\draw [line, thick, dotted] (26) to [out=40,in=140] (S2);
\draw [line, thick, dotted] (5) to [out=30,in=160] (6); 
\draw [line, thick, dotted] (6) to [out=30,in=160] (S);
\draw [line, thick, dotted] (5) to [out=150,in=30] (4); 
\draw [line, thick, dotted] (4) to [out=140,in=35] (2); 
\draw [line, thick, dotted] (2) to [out=150,in=25] (1); 
\draw [line, thick, dotted] (1) to [out=150,in=25] (S0); 	 	
\draw [line, thick, dashed] (P0) to [out=0,in=240] (1);
\draw [line, thick, dashed] (1) to [out=300,in=60] (11); 
\draw [line, thick, dashed] (11) to [out=300,in=60] (21);
\draw [line, thick, dashed] (21) to [out=330,in=210] (22); 
\draw [line, thick, dashed] (11) to [out=330,in=210] (12);
\draw [line, thick, dashed] (12) to [out=330,in=210] (14);
\draw [line, thick, dashed] (14) to [out=120,in=160] (S1);		
\draw [line, thick, dashed] (22) to [out=335,in=205] (24);
\draw [line, thick, dashed] (24) to [out=335,in=205] (25);
\draw [line, thick, dashed] (25) to [out=335,in=205] (26);
\draw [line, thick, dashed] (26) to [out=335,in=205] (S2);
\draw [line, thick] (26) to [out=15,in=165] (S2);
\draw [line, thick] (21) to [out=15,in=165] (22);
\draw [line, thick] (P2) to [out=15,in=165] (21);
\draw [line, thick] (25) to [out=15,in=165] (26);
\draw [line, thick] (24) to [out=15,in=165] (25);
\draw [line, thick] (22) to [out=30,in=150] (24);

\draw  (S0) -- (1) (S) -- (6); 
\draw  (S1) -- (14); 
\draw (S2) -- (26) (P2) -- (21) (P0) -- (1) (P1) -- (15); 
\node[draw, circle, scale=1] (lbroker) at (-1,3) 			  {};
\node[text width=1.5cm] at (0, 3) {\scriptsize \textit{Broker}};
\node[draw, rectangle, scale=1] (lpub)	at  (-1, 2.6) 		{};
\node[text width=3cm] at (0.7, 2.6) {\scriptsize \textit{Publisher and Subscriber}};		
\path [line, dashed, thick] (1, 3) -- (1.5, 3);
\node[text width=3.1cm] at (3.2, 3) {\scriptsize \textit{Publications from $P_{0}$}};	
\path [line, dotted, thick] (4.8, 3) -- (5.3, 3);
\node[text width=3.1cm] at (7, 3) {\scriptsize \textit{Publications from $P_{1}$}};	
\path [line, thick] (3, 2.6) -- (3.5, 2.6);
\node[text width=3.1cm] at (5.2, 2.6) {\scriptsize \textit{Publications from $P_{2}$}};
\end{tikzpicture}
\captionof{figure}{Static publication routing in clustered SCOT. The sets of interested subscribers for $P_{0}$, $P_{1}$, and $P_{2}$ are $\{S_{1}, S_{2}\}$, $\{S, S_{0}, S_{1}, S_{2}\}$, and $\{S_{2}\}$ respectively.}
In Algorithm 3, getDistinctSubs returns a next destination based distinct list of subscriptions to route publication \textit{n}. The host broker forwards \textit{n} to next destination links in the same cluster (lines 2-6), and to secondary brokers of TSCs (lines 7-12).
\begin {algorithm}
\footnotesize
\KwIn{$n:$ a publication message\;}
\KwOut{$DL:$ a list of next destinations that should receive $n$ \;}
\If{$isHostBroker (n.sender)$}{
	$IS \gets getDistinctSubs(n)$\;
	\BlankLine
	
	\textit{/* intra--cluster routing of n in host cluster. */}\;
	\ForEach {$s \in IS$}{
		$n.next \gets s.lastHop$\;
		$DL.add(n)$\;		
	}
	
	$CIV^i_{a} \gets getCIV^i_{a}(n)$\;
	
	\BlankLine
	
	\textit{/* inter--cluster routing of n to TSCs. */}\;
	\ForEach {$bit \in CIV^i_{a}$}{
		\If{ $bit$ =  1 $\wedge$  $i \neq j$}{
			$n.next \gets get\_iLink(bit)$\;
			$DL.add(n)$\;
		}
	}
}\Else{
$IS \gets getDistinctSubs(n)$\;
\BlankLine

\textit{/* intra--cluster routing of n to TSCc. */}\;

\ForEach {$s \in IS$}{
	$n.next \gets s.lastHop$\;
	$DL.add(n)$\;		
}
}
\caption{$scotSPR(n)$}
\end{algorithm}	
\subsection{Inter--cluster Dynamic Routing}
\textit{Dynamic Routing} refers to the capability of a pub/sub system to alter routing path in response to congested or failed links and/or congested or failed brokers. Multiple techniques have been proposed to handle dynamic routing in address--based networks where routing paths are calculated from a global view of a network topology graph that is saved on every network router \cite{routing_book}. IP addresses may be used to form clusters in a network area where same network mask can be used which requires a single table entry. However, these techniques are not applicable in pub/sub systems because brokers are aware of only their direct neighbours and destinations addressing is based on contents (i.e., subscriptions). Therefore dynamic routing decisions are decentralized and have to be made without having a global view of an overlay. This section describes \textit{Inter--cluster Dynamic Routing (IDR)} algorithm. It deals with routing when one or more iLinks are congested and publications start queuing up, while discussion about dynamic routing when one or more brokers fail is out of scope of this paper. 

\texttt{OctopiA} leverage the \textit{structuredness} of SCOT to realize subgrouping and eliminate tight coupling to achieve inter--cluster dynamic routing when one or more iLinks are congested. Our approach is scalable because it does not require updates in routing tables and can reduce delivery delays when a large number of publications start accumulating in the output queues. SPR algorithm adds exclusive copies of a publication in the output queues of the target links. If a publisher generates $\gamma$ number of publications in $t_{w}$ interval, and there are $\alpha$ number of target aLinks and $\beta$ target iLinks, the host broker of the publisher enqueue $(\alpha + \beta).\gamma$ number of copies of a publication in $t_{w}$ interval. A High Rate Publisher (HRP) with a high value of $\gamma$ can overwhelm brokers when SPR algorithm is used. IDR algorithm alleviates overwhelmed brokers and dynamically \textit{tries} to find uncongested iLinks to forward publications to TSCs. As parallel iLinks are available in a clustered SCOT, one can be selected without making updates in routing tables. The selected iLink may not be part of a subscription--tree to the TSC. IDR algorithm adds at most \textit{one copy} of a publication, when there are multiple congested output queues of the target links at a broker. For the remaining TSCs, which are not forwarded the publication due to congested output queues, the algorithm sets their index bits in $CIV^i_{p}$ to 1. $CIV^i_{p}$ is then added in the header of the publication that is added in the output queue. The publication that carries $CIV^i_{p}$ is identified as $ CIV^i_{p} - Dynamic$ or $CIV^i_{p} - D$. Using this technique, IDR algorithm keeps the length of the congested output queues of an overwhelmed broker to a minimum, and the load of forwarding the publication to the remaining TSCs, with help of their index bits in $CIV^i_{p}$, is shifted to other brokers using the heuristic that uncongested target iLinks are available down the routing path. \texttt{OctopiA} uses Eq. 3 to find whether an output queue is congested and IDR algorithm should be activated.
\begin{equation}
(Q_{\ell}) . (1+Q_{in}, 1+Q_{out})_{t_{w}}  > \tau 
\end{equation}
$Q_{in}$ and $Q_{out}$ are the number of publications that enter into or leave the output queue in the time window $t_{w}$, respectively. The term $(1+Q_{in}, 1+Q_{out})_{t_{w}}$ is the ratio of $(1+Q_{in})$ to $(1+Q_{out})$, and is known as the \textit{Congestion Element (CE)}. $CE > 1$ indicates that the congestion is increased, while $CE < 1$ shows that congestion is decreased in the last $t_{w}$ interval. An output queue is congestion--free when $CE$ is 1 and the queue length $Q_{\ell}$ is 0. \texttt{OctopiA} saves the values of $Q_{in}$ and $Q_{out}$ in a \textit{Link Status Table (LST)} on each broker, and the values are updated after each $t_{w}$ interval. Dynamic routing with IDR is further explained with the help of three selected cases in the following.\\
\textbf{Case I -- Overloaded iLinks: } This case describes dynamic routing when one or more (but not all) target iLinks are congested. The relation: $\big ( 1 \leq |iLink|_{OL} < \beta \big )$, where $|iLink|_{OL}$ is the number of congested target iLinks at a broker, expresses that one or more target iLinks are congested. After detecting the congested output queues, dynamic routing is activated by forwarding $CIV^i_{p} - D$, where $i \in V_{cf}$ identifies index of the broker (or cluster) which detected congestion, to the other brokers down the dynamic routing path. No congested output queue of the target iLinks receives the publication. $CIV^i_{p}$ is added (with $CIV^i_{p} - D$) into the output queue of an uncongested target iLink, and unlike to other two contexts, is not saved in any routing tables. If there are multiple uncongested target iLinks, $CIV^i_{p}-D$ is added into the output queue with the least value of $Q_{\ell}$. The number of publications added in $t_{w}$ interval is $(\alpha + \theta ).\gamma$, where $\theta$ is the number of uncongested target iLinks, and $\theta < \beta$. Fig. 7(b) shows that after detecting the congestion using Eq. 3, \textit{B(b, 3)} switches to IDR algorithm with the two candidate iLinks, $l \langle(b,3), (b,2) \rangle$, and $l \langle(b,3), (b,1) \rangle$, that can receive $CIV^3_{p}-D$. Presumably, having the least value of $Q_{\ell}$, the link $l \langle(b,3), (b,1) \rangle$ is selected to route $CIV^3_{p}-D$. \textit{B(b, 3)} sets the bit $CIV^3_{p}[0]$ to 1 (and all other bits to 0), and adds $CIV^3_{p}-D$ in the output queue of $l \langle(b,3), (b,1) \rangle$. The bit pattern \textit{0001} of $CIV^3_{p}$ indicates that $C_{0}$ is a TSC. As the target iLink $l \langle(b,1), (b,0) \rangle$ is also congested, the $CIV^3_{p}-D$ is forwarded to \textit{B(c, 1)} using the subscription--tree of \textit{S2}. At \textit{B(c, 1)}, the publication is eventually forwarded to broker \textit{B(c, 0)} of $C_{0}$ as the target iLink $l \langle(c, 1), (c, 0) \rangle$ is uncongested. Since $C_{0}$ was the only TSC, \textit{B(c, 1)} removes $CIV^3_{p}$ from $CIV^3_{p} - D$ and routing switches to static mode. To forward a publication to all subscribers, SPR algorithm generates 8 IMs, while IDR algorithm generated 7 IMs. The dynamic routing path generated by IDR algorithm for \textit{S1} contained one extra broker but no congested iLink.\\
\textbf{Case II -- All iLinks Overloaded: } The condition $\big (|iLink|_{OL} = \beta \big)$ at a broker indicates that all the target iLinks are congested. We assume that at least one target aLink is uncongested and can be selected to forward $CIV^i_{p}-D$. If more than one aLinks are available, the one with least value of $Q_{\ell}$ is selected. The number of publications added to the output queues is $\alpha .\gamma$. Fig. 7(c) illustrates that $l \langle(b,3),(b,0)\rangle$ and $l \langle(b,3),(b,2)\rangle$ are congested and alternative paths to forward the publication to TSCs $C_{0}$ and $C_{2}$ should be searched. The link $l \langle(b,3),(b,1)\rangle$ is not useful because it forwards publications to a non--TSC cluster $C_{1}$ (Case III indicates that this generates loops). Presumably, having least value of $Q_{\ell}$, $CIV^3_{p}-D$ (with $CIV^3_{p}$ \textit{0101}) is added into the output queue of $l \langle(b,3),(c,3)\rangle$. \textit{B(c, 3)} forwards the publication onto two uncongested target iLinks to TSCs $C_{0}$ and $C_{2}$ after removing $CIV^3_{p}$. IDR generated 6 IMs and eliminated two congested target iLinks when sending a publication to \textit{S1}, \textit{S2}, and \textit{S5}, while SPR algorithm generates 7 IMs.\\
\textbf{Case III -- All Links Overloaded:} The condition $\big( |iLinks|_{OL} + |aLinks|_{OL} = \alpha + \beta \big)$, where $|aLinks|_{OL}$ is the number of congested target aLinks at a broker, indicates that no uncongested target link is available. $CIV^i_{p}-D$ is added into the congested output queue of the target iLink with least value of $Q_{\ell}$. The congested target aLink, even if has the least value of $Q_{\ell}$, is not selected because of the possibility of having more congested aLinks down the routing path in the same cluster. The $(1+\alpha)$ number of copies of the publication in generated in $t_{w}$ interval. Fig. 7(d) shows that $CIV^3_{p}-D$ ($CIV^3_{p}$ is \textit{0011}) is forwarded onto the target iLink $l \langle(b,3),(b,2)\rangle$, which presumably has the least value of $Q_{\ell}$. At \textit{B(b, 2)}, the target link $l \langle(b,2),(b,1)\rangle$ is also congested, while the target iLink $l \langle(b,2),(b,0)\rangle$, is uncongested. \textit{B(b, 2)} sets $CIV^3_{p}[0]$ to 0 and forwards $CIV^3_{p}-D$ (with $CIV^3_{p}$ \textit{0010}) to \textit{B(b, 0)}. $CIV^3_{p}$ indicates that the publication should be forwarded to $C_{1}$. However, \textit{B(b, 0)} forwards $CIV^3_{p}-D$ to \textit{B(a, 0)} because the target iLink $l \langle(b,0),(b,1)\rangle$ is congested and a search for uncongested target iLink is still possible at \textit{B(a, 0)} because the link $l \langle(b,0),(a,0)\rangle$ is part of the subscription--tree of $S1$, which is not sent the publication yet. At \textit{B(a, 0)}, the only available target iLink $l \langle(a,0),(a,1)\rangle$ is also congested and there is no more aLink available in $C_{0}$ cluster (the publication cannot be sent back to B(b, 0) to avoid message loops). As a last option, \textit{B(a, 0)} forwards the publication to $B(a, 1) \in C_{1}$ after removing $CIV^3_{p}$ from the header. IDR algorithm generated 8 IMs and could eliminate only one congested iLinks, while SPR algorithm generates 7 IMs to forward a publication. Dynamic routing paths for \textit{S1} and \textit{S2} had 3 additional brokers. To prevent message loop, cluster index bit of a TSC in $CIV^i_{p}$ is set to 0 after the publication is delivered to any broker of the TSC. For example, in order to avoid the congested target iLink $\langle(a,0),(a,1)\rangle$, a loop is possible between \textit{B(b, 2), B(b, 0), B(a, 0)} and \textit{B(a, 2)} if the cluster index bits are not set to 0.

\begin {algorithm}
\footnotesize
\KwIn{$n:$ a publication message\;}
\KwOut{$DL:$ a list of next destinations that should receive $n$ \;}
\BlankLine

\textit{/* get distinct interested subscribers from this cluster */}\;	
$IS \gets getDistinctSubs(n)$\;
\ForEach {$s \in IS$}{
	$n.next \gets s.lastHop$\;
	$DL.add(n)$\;		
}

\BlankLine

\textit{/* prepare $CIV^i_{p}$ at the host broker of a publisher */}\;
\If{$isHostBroker (n.sender)$}{	
	$CIV^i_{a} \gets getCIV^i_{a}(n)$\;			
	\BlankLine
	\ForEach {$bit \in CIV^i_{a}$}{
		\If{ $bit$ = 1}{
			\BlankLine
			
			\textit{/* set the $CIV^i_{p}[j]$ for overloaded link */}\;
			\If{$iLinkOverloaded(j, \beta)$ = true}{
				$n.CIV^i_{p}[j] \gets 1$\;
			}\Else{
			$n.next \gets get\_iLink(j)$\;
			$DL.add(n)$\;	
		}
	}
}
} \Else{
$CIV^i_{p} \gets n.CIV^i_{p}$\;
$n.CIV^i_{p} \gets \emptyset$\;
\ForEach {$bit \in CIV^i_{p}$}{
	\If{ $bit = 1$ $ \wedge$  $iLinkOverloaded(j, \beta) = false$}{
		$n.next \gets get\_iLink(j)$\;
		$DL.add(n)$\;
		$CIV^i_{p}[j] \gets 0$\;
	}
}

$n.CIV^i_{p} \gets CIV^i_{p}$\;
}

\BlankLine
\textit{/* if there are some TSCs, which should receive n */}\;
\If{$n.CIV^i_{p} \not= 0$}{
	$\mu1 \gets getLeastLoaded\_aLink(\alpha)$\;
	$\mu2 \gets getLeastLoaded\_iLink(\beta)$\;
	\BlankLine
	\textit{/* Case I... */}\;
	\If{$isOverloaded(\mu2) = false$}{
		$adjustCIV^i_{p}-D(DL,  \mu2)$\;
	}
	\textit{/* Case II... */}\;
	\ElseIf{$isOverloaded(\mu1) = false$}{
		$adjustCIV^i_{p}-D(DL,  \mu1)$\;
	}
	\textit{/* Case III... */}\;
	\Else{
		$n.next \gets \mu1$\;
		$n.CIV^i_{p}[j] \gets 0$\;
		$DL.add(n)$\;
	}
}

\BlankLine		
\label{algo:idr}
\caption{$scotIDR(n)$}
\end{algorithm}

In Algorithm 4, the \textit{getDistinctSubs} method follows the principle of downstream replication and all target aLinks (whether overloaded) receive a copy of publication \textit{n} (lines 2-5). The host broker of a publisher initially adds exclusive copies of \textit{n} for next destinations to TSCs and sets the cluster index bits for overloaded iLinks (lines 7-16). A broker checks for the unoverloaded target iLink to forwarded a copy of \textit{n} to a TSC (lines 18-24). The method \textit{iLinkOverloaded} takes current cluster index \textit{j} and list of all target iLinks $\beta$ to check if an iLink is overloaded. The method \textit{get\_iLink} takes index of a TSC to provide an iLink to a broker of that TSC. $\mu1$ and $\mu2$ represent least loaded aLink and iLink respectively. The method $adjustCIV^i_{p}-D$ prepares the $CIV^i_{p}-D$ publication by adding $CIV^i_{p}$ in the header of a copy of \textit{n}. Three cases are handleled based on overloaded and unoverloaded iLinks and aLinks (lines 27-40).

IDR algorithm has two limitations: (i) no support for intra--cluster dynamic routing, and (ii) a subscription--tree of an interested subscriber is required in a cluster to search uncongested target links. For example, in Fig. 7(d), the publication has to be forwarded onto congested links at \textit{B(b, 3)} although $l \langle(b,3),(a,3)\rangle$ is uncongested. Since $l \langle(b,3),(a,3)\rangle$ is not included in subscription--trees of \textit{S4} and \textit{S3}, it cannot be used. An extended version of IDR algorithm to handle these limitations is part of the future work.
\section{Evaluation}
We implemented SPR and IDR algorithms in \texttt{OctopiA}, developed on top of an open source pub/sub tool PADRES \cite{PADRESBookChapte}. For a comparison with state--of--the--art, we used TID--based routing algorithms \cite{Li_ADAP}.\\
\textbf{Setup ---}
Fig. 9 shows the factors of SCOT topology $\mathbb{S}_{e}$, which we used for evaluation and comparison of SPR and IDR algorithms with \textit{TID--based static (TID--S)} and \textit{TID--based dynamic (TID--D)} routing algorithms. The $G_{af}$ factor of $\mathbb{S}_{e}$ is an acyclic topology of 14 brokers, with 4 inner  (\textit{F, G, H} and \textit{I}), and 10 edge brokers (\textit{A, B, C, D, E, J, K, L, M} and \textit{N}). $\mathbb{S}_{e}$ had 16 inner and 40 edge brokers for a total of 54 brokers and was deployed on a cluster of 7 computing nodes connected through one dedicated switch. Each node had 2 Intel Xeon E5--2620 CPU with total 12 cores of 2.1 GHz each, and 64 GB RAM.\\  
Stock datasets are commonly used to generate workloads for evaluations of pub/sub systems \cite{agg15}. We used a dataset of 500 stock symbols from Yahoo Finance!, where each stock publication had 7 distinct attributes. Advertisements and subscriptions were generated synthetically with 7 distinct attributes and 2\% selectivity. This high dimension data require high computation for matching and routing of publications. We randomly distributed publishers and subscribers where each one of them either issued one advertisement or subscription.\\
\begin{tikzpicture}[thick,scale=1.2]
\scriptsize
\def\scaleSCOT {1}
\def\y {6}
\def\x {0}
\def\xInc {0.7}
\def\yInc {0.7}

\node[text width=3cm] at (\x + 0.5,\y-\yInc) {\small $\mathbb{S}_{e} : $};

\node[draw, circle, scale=\scaleSCOT] (A) at (\x + \xInc*1,\y) {$A$};
\node[draw, circle, scale=\scaleSCOT] (B) at (\x + \xInc*2,\y) {$B$};
\node[draw, circle, scale=\scaleSCOT] (C) at (\x + \xInc*3,\y) {$C$};
\node[draw, circle, scale=\scaleSCOT] (D) at (\x + \xInc*4,\y) {$D$};

\node[draw, circle, scale=\scaleSCOT] (E) at (\x ,\y-\yInc) {$E$};
\node[draw, circle, scale=\scaleSCOT] (F) at (\x + \xInc*1,\y-\yInc) {$F$};
\node[draw, circle, scale=\scaleSCOT] (G) at (\x + \xInc*2,\y-\yInc) {$G$};
\node[draw, circle, scale=\scaleSCOT] (H) at (\x + \xInc*3,\y-\yInc) {$H$};
\node[draw, circle, scale=\scaleSCOT+0.2] (I) at (\x + \xInc*4,\y-\yInc) {$I$};
\node[draw, circle, scale=\scaleSCOT] (J) at (\x + \xInc*5,\y-\yInc) {$J$};

\node[draw, circle, scale=\scaleSCOT] (K) at (\x + \xInc,\y-\yInc*2) {$K$};
\node[draw, circle, scale=\scaleSCOT] (L) at (\x + \xInc*2,\y-\yInc*2) {$L$};
\node[draw, circle, scale=\scaleSCOT] (M) at (\x + \xInc*3,\y-\yInc*2) {$M$};
\node[draw, circle, scale=\scaleSCOT] (N) at (\x + \xInc*4,\y-\yInc*2) {$N$};

\draw [thick] (A) -- (F) (B) -- (G) (C) -- (H) (D) -- (I) (E) -- (F) (F) -- (G) ;
\draw [thick] (G) -- (H)  (I) -- (J) (K) -- (F) (L) -- (G) (M) -- (H) (N) -- (I) (I) -- (H);
\node[draw, rectangle, minimum size=0.3cm] (op)    at (4.2,\y-\yInc) {};
\node[draw, circle, scale=\scaleSCOT+0.1] (vc) at (5.3,\y-\yInc+0.7) {$0$};
\node[draw, circle, scale=\scaleSCOT+0.1] (va) at (6,\y-\yInc*2+0.2) {$1$};
\node[draw, circle, scale=\scaleSCOT+0.1] (vb) at (4.7,\y-\yInc*2+0.2) {$2$};
\node[draw, circle, scale=\scaleSCOT+0.1] (vd) at (5.3,\y-\yInc-0.05) {$3$};		
\draw [dashed, thick] (va) -- (vb) (vb) -- (vc) (vc) -- (va) (vd) -- (va) (vd) -- (vb) (vd) -- (vc);
\end{tikzpicture}

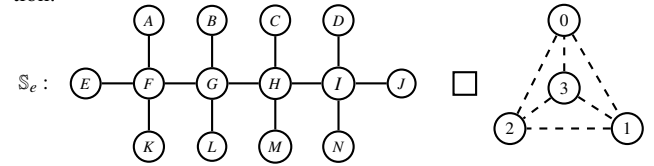
\captionof{figure}{CPUG operands of topology  $\mathbb{S}_{e}$.}
\label{fig:evalTop}
\textbf{Metrics ---}
To demonstrate the effectiveness, experiments were run in an unclustered (for PADRES) and clustered (for \texttt{OctopiA}) $\mathbb{S}_{e}$. We used the following \textit{metrics}.\\ 
\textit{Advertisement delay} is the maximum time elapsed by an advertisement to reach every broker of a cyclic overlay. In \texttt{OctopiA}, advertisement delay is expected to be less because an advertisement is broadcast to brokers in a region and generates advertisement--tree of length 1.\\ 
In traditional pub/sub systems, \textit{Subscription delay} is the maximum time elapsed as a subscription is forwarded from the subscriber to brokers with matching advertisements. \texttt{OctopiA} performs cluster--level subscription broadcast, without considering matching advertisement. The study of difference between subscription delays in two cases is worthwhile.\\
\textit{Publication delay} measures end--to--end latency of a publication delivery and is an important metric.\\
\textit{Matching delay} is the time taken by a broker to find matching subscriptions. As number of subscriptions saved at each broker in both tools is expected to vary, this metric provides useful information about in--broker processing.\\ 
The number of \textit{Inter--broker Messages (IMs)} depends on the number of overlay hops in routing paths. \texttt{OctopiA} generates routing paths of shortest lengths, therefore, investigating the number of IMs generated in both cases is important.\\
\textit{Size of SRT and CLT} provide the number of advertisements saved in routing tables in both cases. As both tools use different approaches to forward advertisements, knowing difference of these values provide useful insight.\\
\textit{Size of PRT} measures the number of subscriptions saved in PRTs. As both the tools use different SBP algorithms, it is important to investigate the difference between average size of PRTs.

Aggregation techniques like covering can be used to reduce size of CLTs, SRTs and PRTs. These techniques are developed for acyclic overlays and can be used at cluster--level in \texttt{OctopiA}. However, covering is not considered because PADRES does not support covering in cyclic overlays.\\ 
\begin{figure*}
	\centering
	\captionsetup[subfigure]{labelformat=empty}
	\begin{subfigure}[b]{0.18\textwidth}
		\includegraphics[trim=1cm 0.5cm 0.5cm 1.5cm, width=0.9\textwidth]{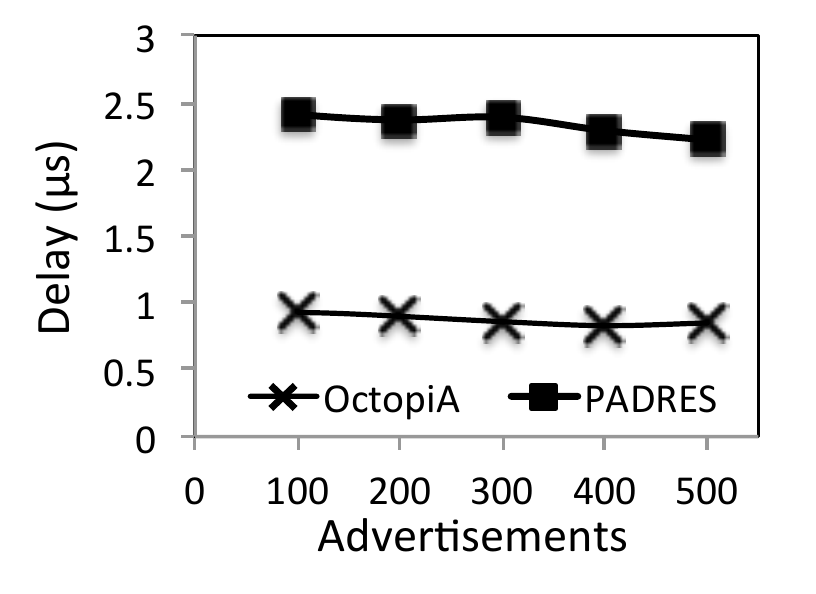}
		\subcaption{\scriptsize Fig. 7: Advertisement Delay.}
		\label{fig:S1}
	\end{subfigure}
	~	
	\begin{subfigure}[b]{0.18\textwidth}
		\includegraphics[trim=1cm 0.5cm 0.5cm 2cm, width=0.9\textwidth]{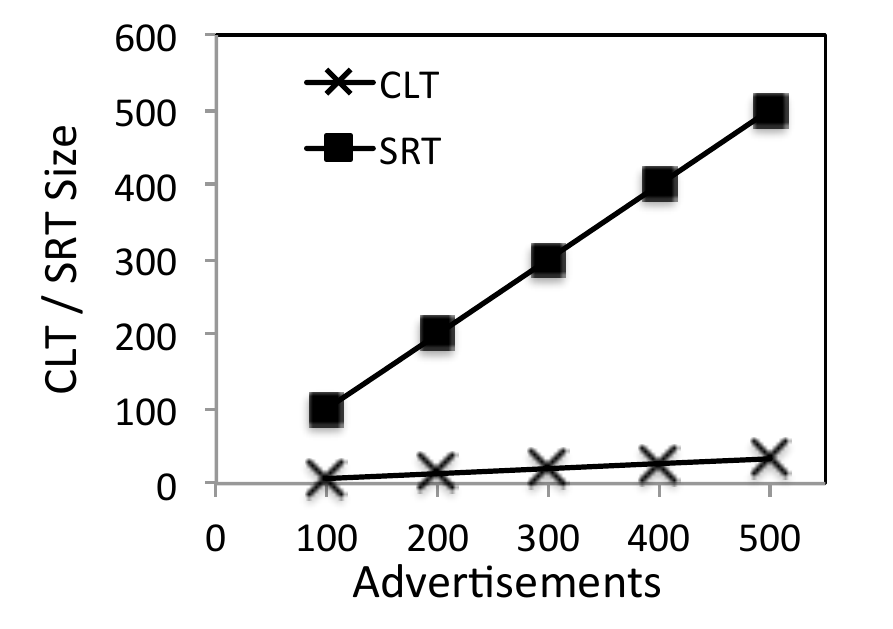}
		\caption{\scriptsize Fig. 8: Size of CLTs/SRTs.}
		\label{fig:S2}
	\end{subfigure}
	~ 
	\begin{subfigure}[b]{0.18\textwidth}
		\includegraphics[trim=1cm 0.5cm 0.5cm 2cm, width=0.9\textwidth]{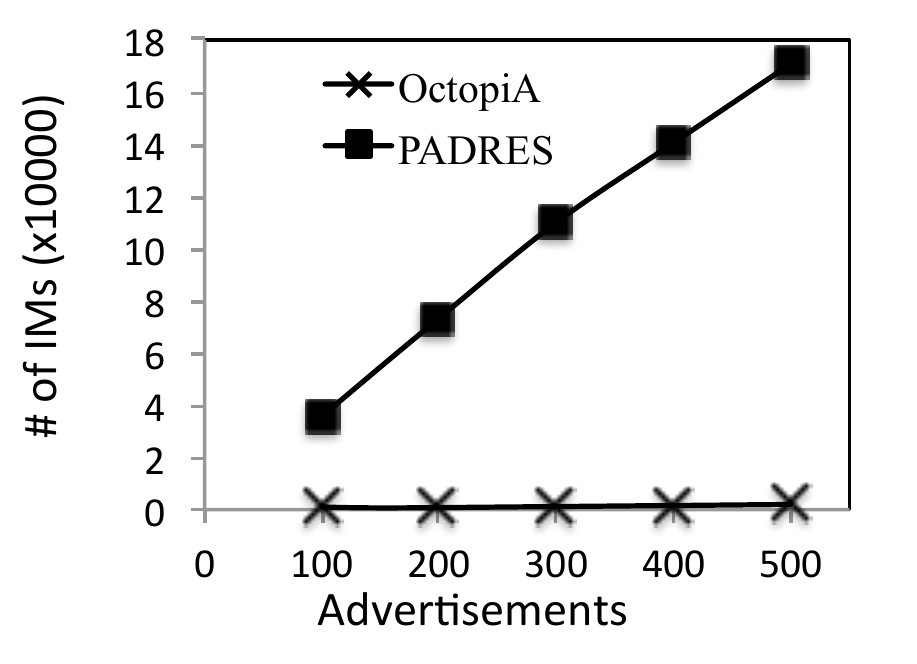}
		\caption{\scriptsize Fig. 9: \# of IMs in ABP.}
		\label{fig:P10}
	\end{subfigure}
	~ 
	\begin{subfigure}[b]{0.17\textwidth}
		\includegraphics[trim=1cm 0.5cm 0.5cm 2cm, width=0.9\textwidth]{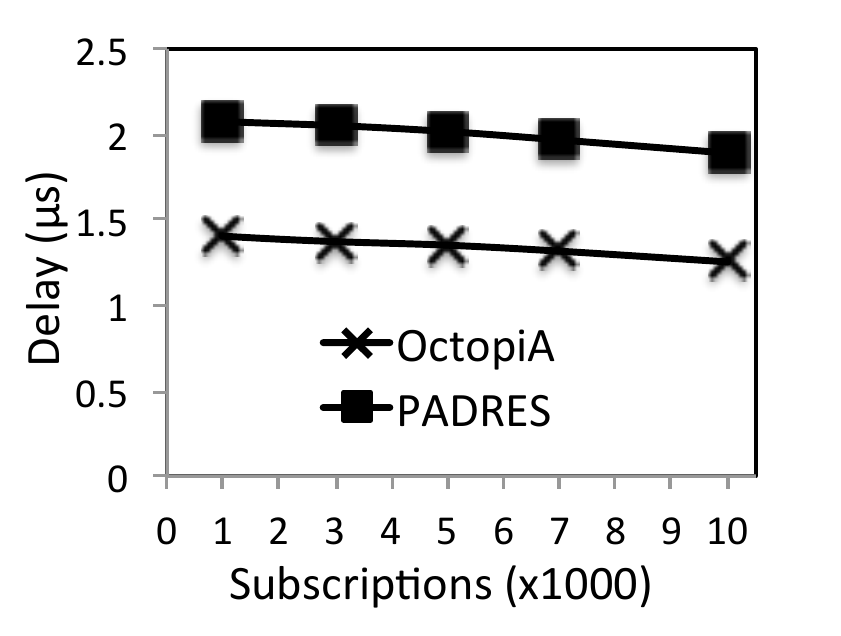}
		\caption{\scriptsize Fig. 10: Subscription delay.}
		\label{fig:pp1}
	\end{subfigure}
	~ 
	\begin{subfigure}[b]{0.19\textwidth}
		\includegraphics[trim=1cm 0.5cm 0.5cm 2cm, width=0.9\textwidth]{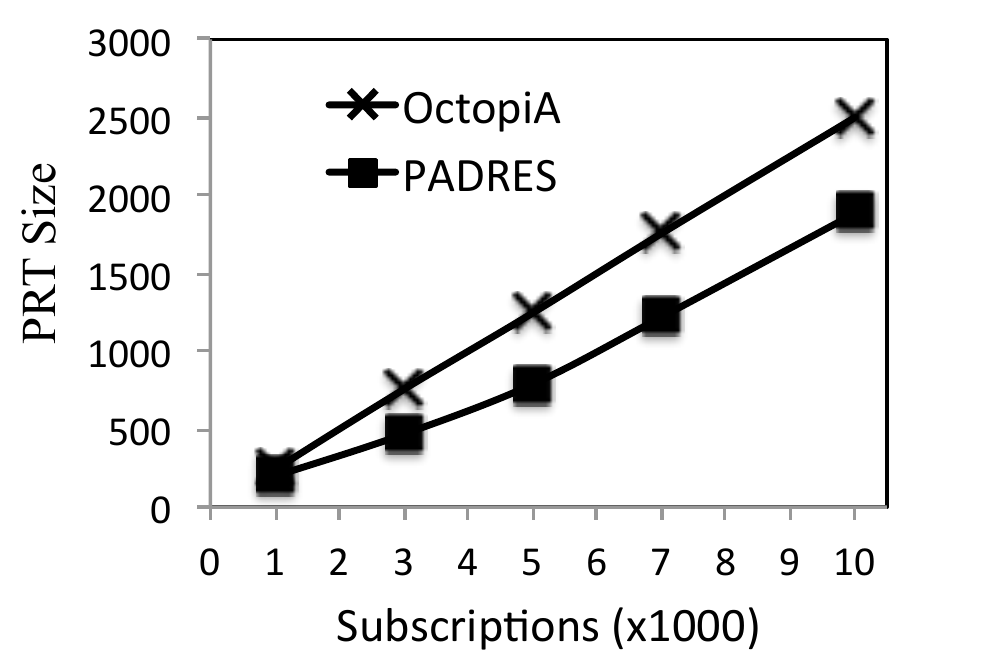}
		\caption{\scriptsize Fig. 11: PRT size.}
		\label{fig:P1}
	\end{subfigure}
\end{figure*}
\begin{figure*}
	\centering
	\captionsetup[subfigure]{labelformat=empty}
	\begin{subfigure}[b]{0.17\textwidth}
		\includegraphics[trim=1cm 0.5cm 0.5cm 1.2cm, width=0.9\textwidth]{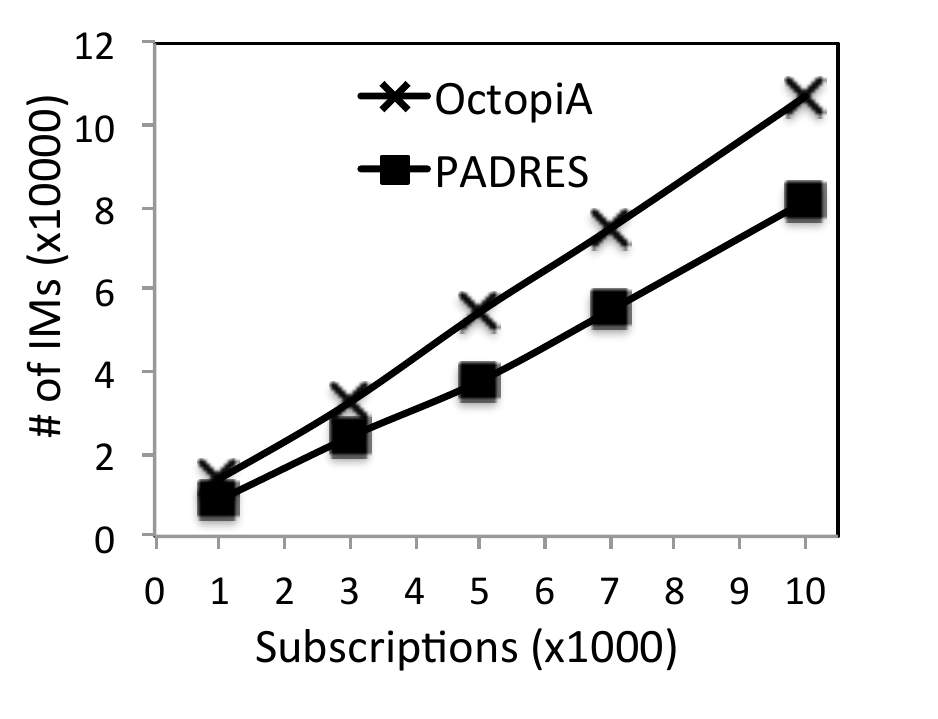}
		\caption{\scriptsize Fig. 12: \# of IMs in SBP}
		\label{fig:P8}
	\end{subfigure}
	~
	\begin{subfigure}[b]{0.18\textwidth}
		\includegraphics[trim=1cm 0.5cm 0.5cm 1.2cm, width=0.9\textwidth]{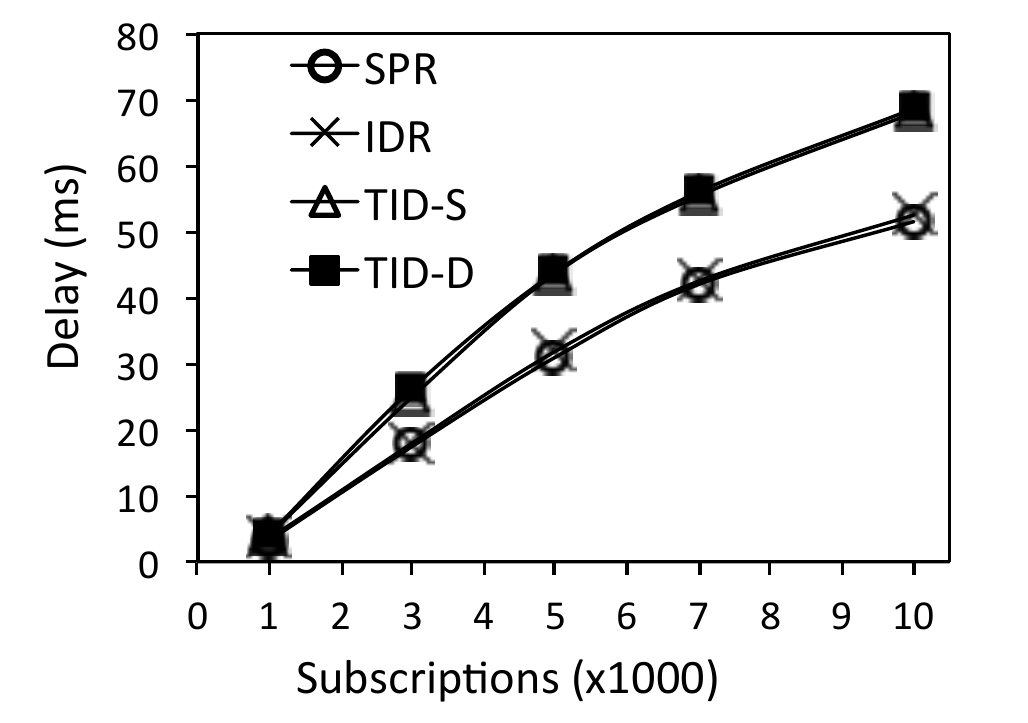}
		\caption{\scriptsize Fig. 13: Publication delay}
		\label{fig:S55}
	\end{subfigure}
	~
	\begin{subfigure}[b]{0.18\textwidth}
		\includegraphics[trim=1cm 0.5cm 0.5cm 1.2cm, width=0.9\textwidth]{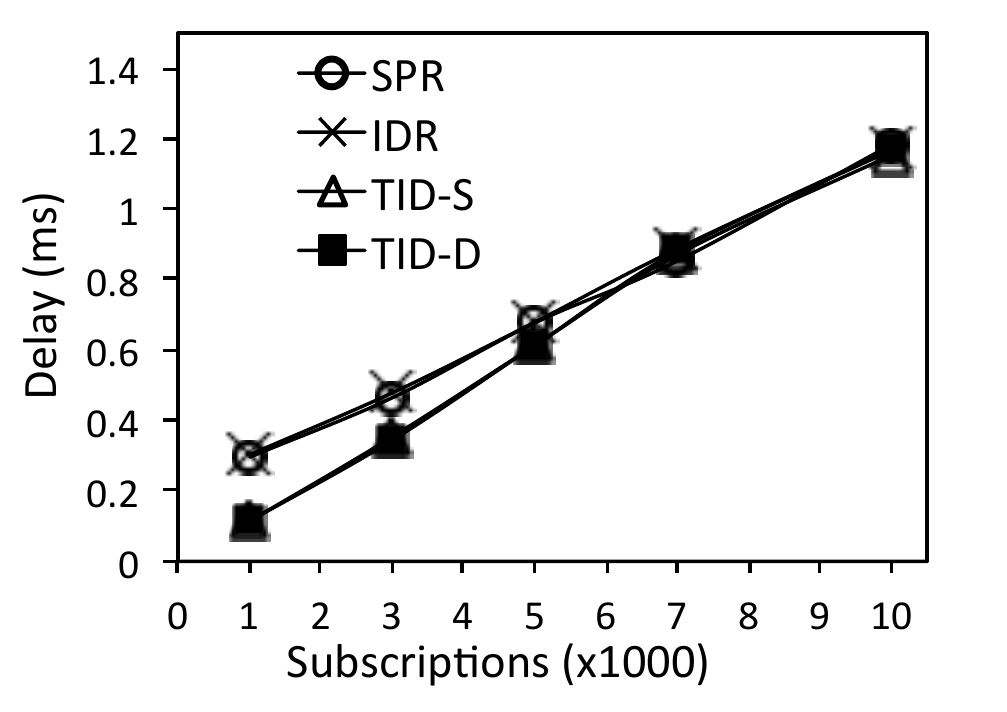}
		\caption{\scriptsize Fig. 14: Matching delay}
		\label{fig:P4}
	\end{subfigure}
	~ 
	\begin{subfigure}[b]{0.17\textwidth}
		\includegraphics[trim=1cm 0.5cm 0.5cm 1.2cm, width=0.9\textwidth]{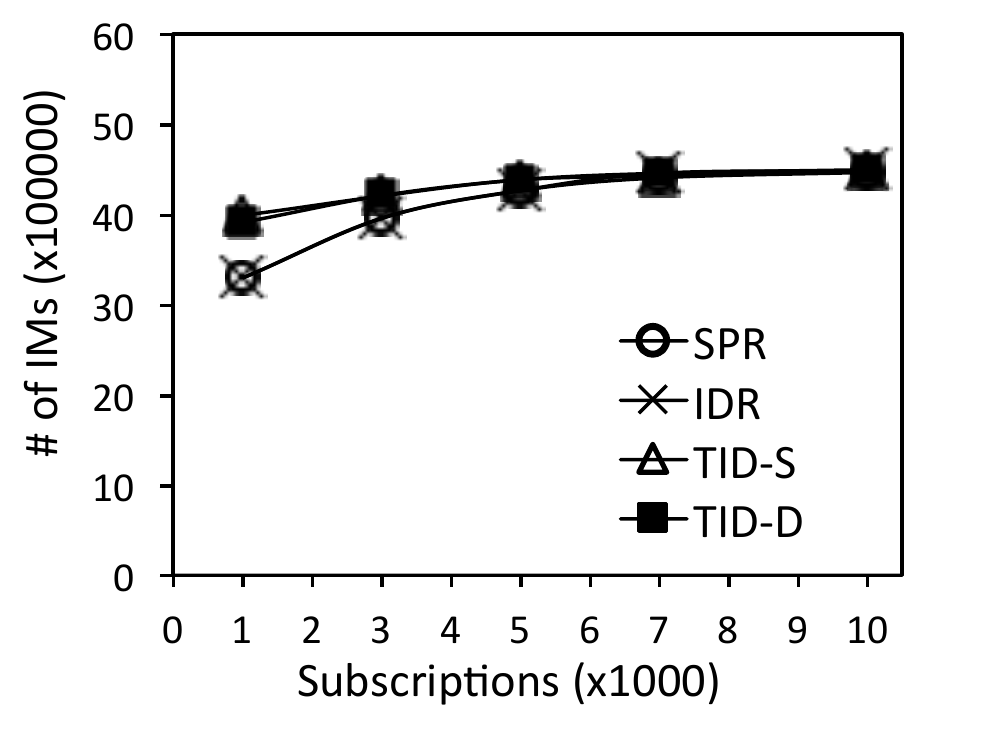}
		\caption{\scriptsize Fig. 15: \# of IMs}
		\label{fig:P5}
	\end{subfigure}
	~
	\begin{subfigure}[b]{0.18\textwidth}
		\includegraphics[trim=1cm 0.5cm 0.5cm 1.2cm, width=0.9\textwidth]{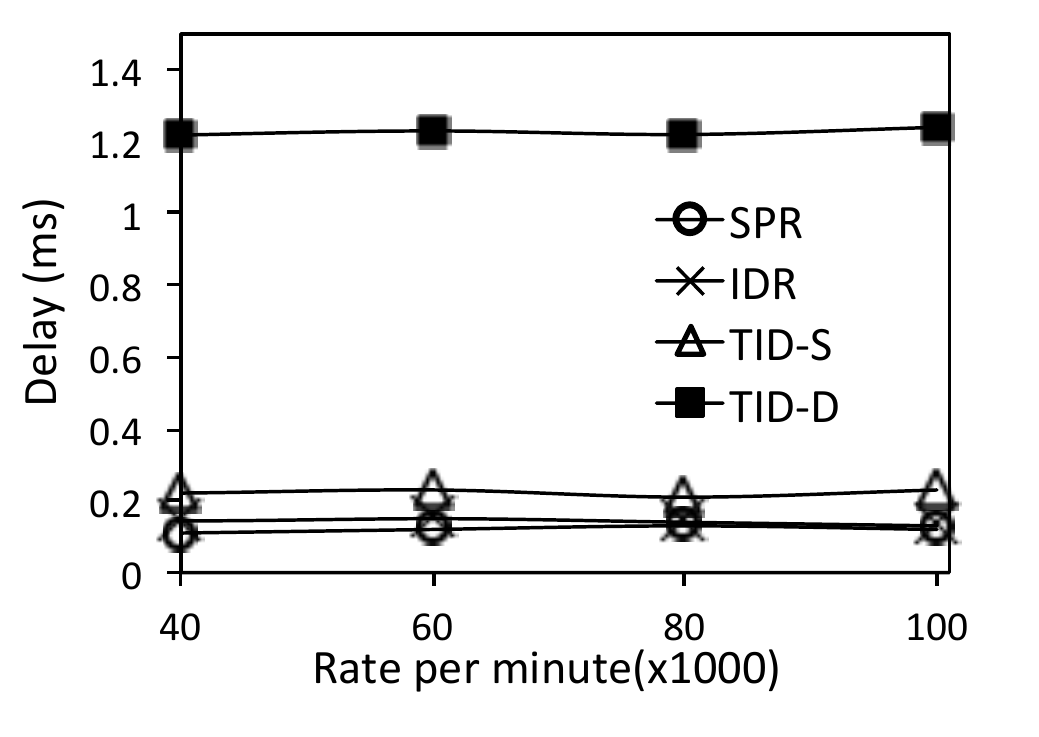}
		\caption{\scriptsize Fig. 16: Path selection delay}
		\label{fig:P9}
	\end{subfigure}		
\end{figure*}
\textbf{Results ---}
The results presented in this section cover fours important aspects of evaluation: (i) \textit{Publisher Scalability}, (ii) \textit{Subscriber Scalability}, (iii) \textit{End to End publication delivery}, and (iv) \textit{Dynamic Routing}.\\
\begin{figure*}
	\centering
	\captionsetup[subfigure]{labelformat=empty}
	\begin{subfigure}[b]{0.22\textwidth}
		\includegraphics[trim=0.5cm 0.5cm 0.5cm 1cm, width=0.95\textwidth]{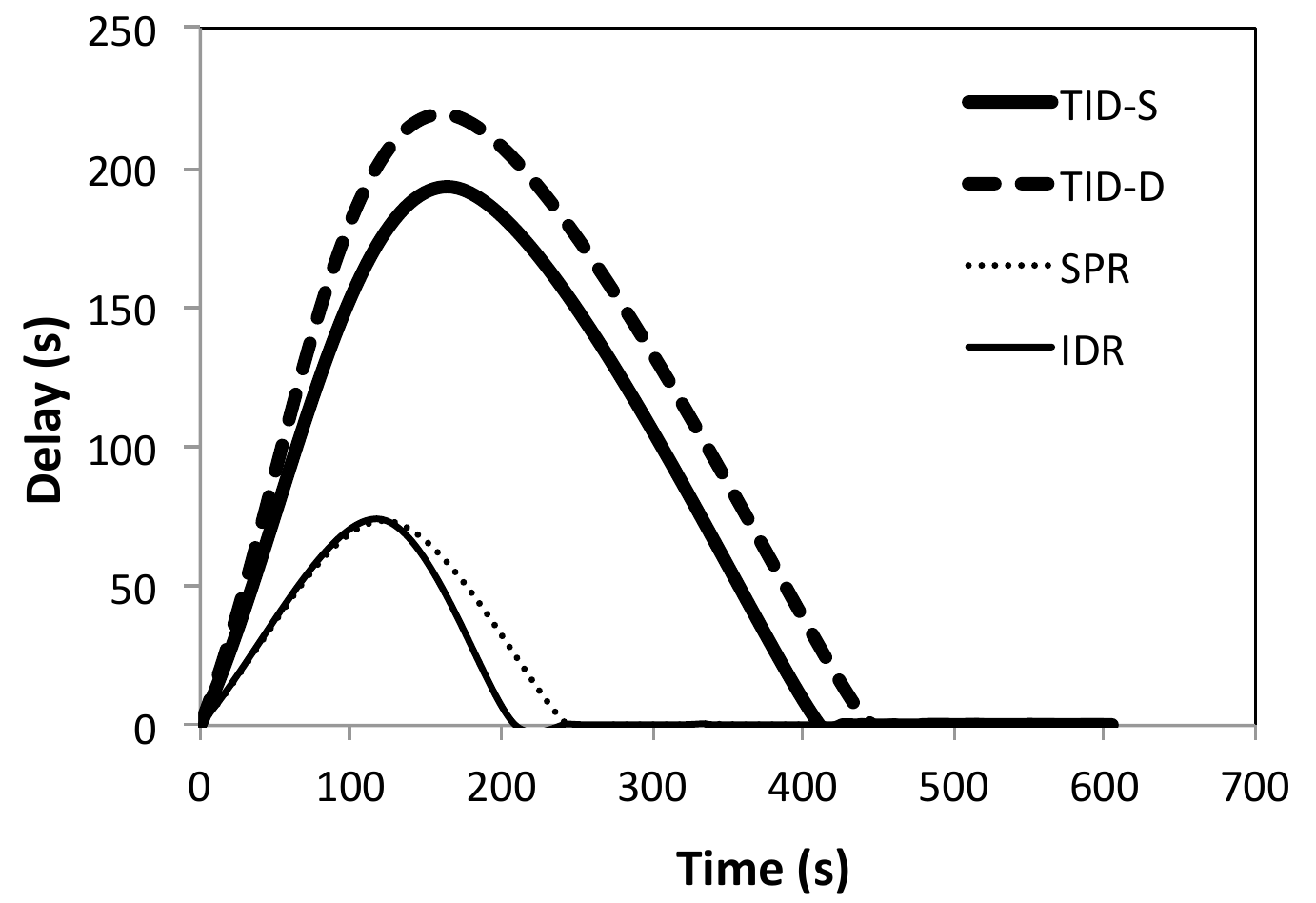}
		\subcaption{\scriptsize Fig. 17: 100K messages per minute}
	\end{subfigure}
	~
	\begin{subfigure}[b]{0.22\textwidth}
		\includegraphics[trim=0.5cm 0.5cm 0.5cm 1cm, width=0.95\textwidth]{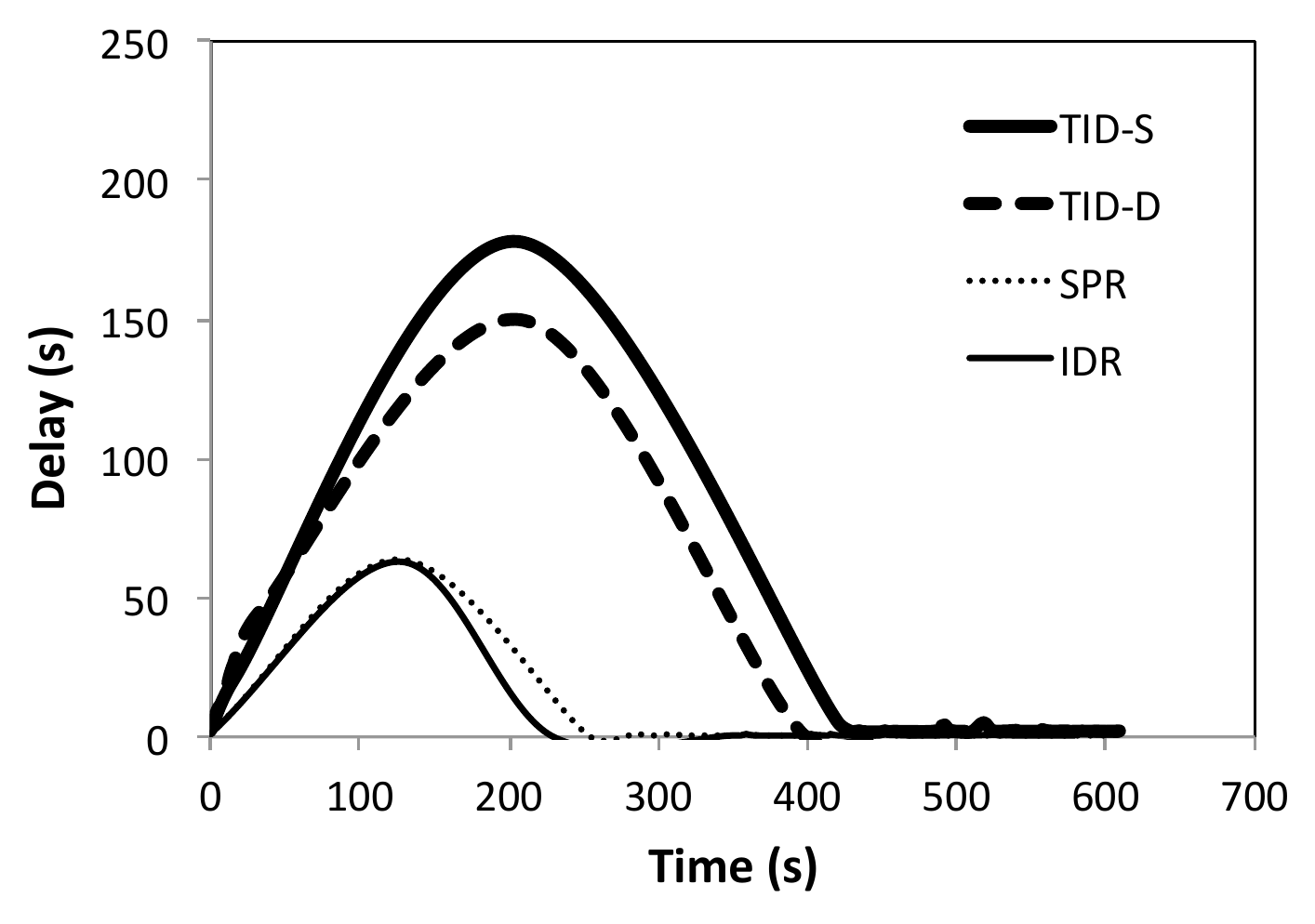}
		\subcaption{\scriptsize Fig. 18: 80K messages per minute}
	\end{subfigure}
	~
	\begin{subfigure}[b]{0.23\textwidth}
		\includegraphics[trim=0.5cm 0.5cm 0.5cm 1cm, width=0.95\textwidth]{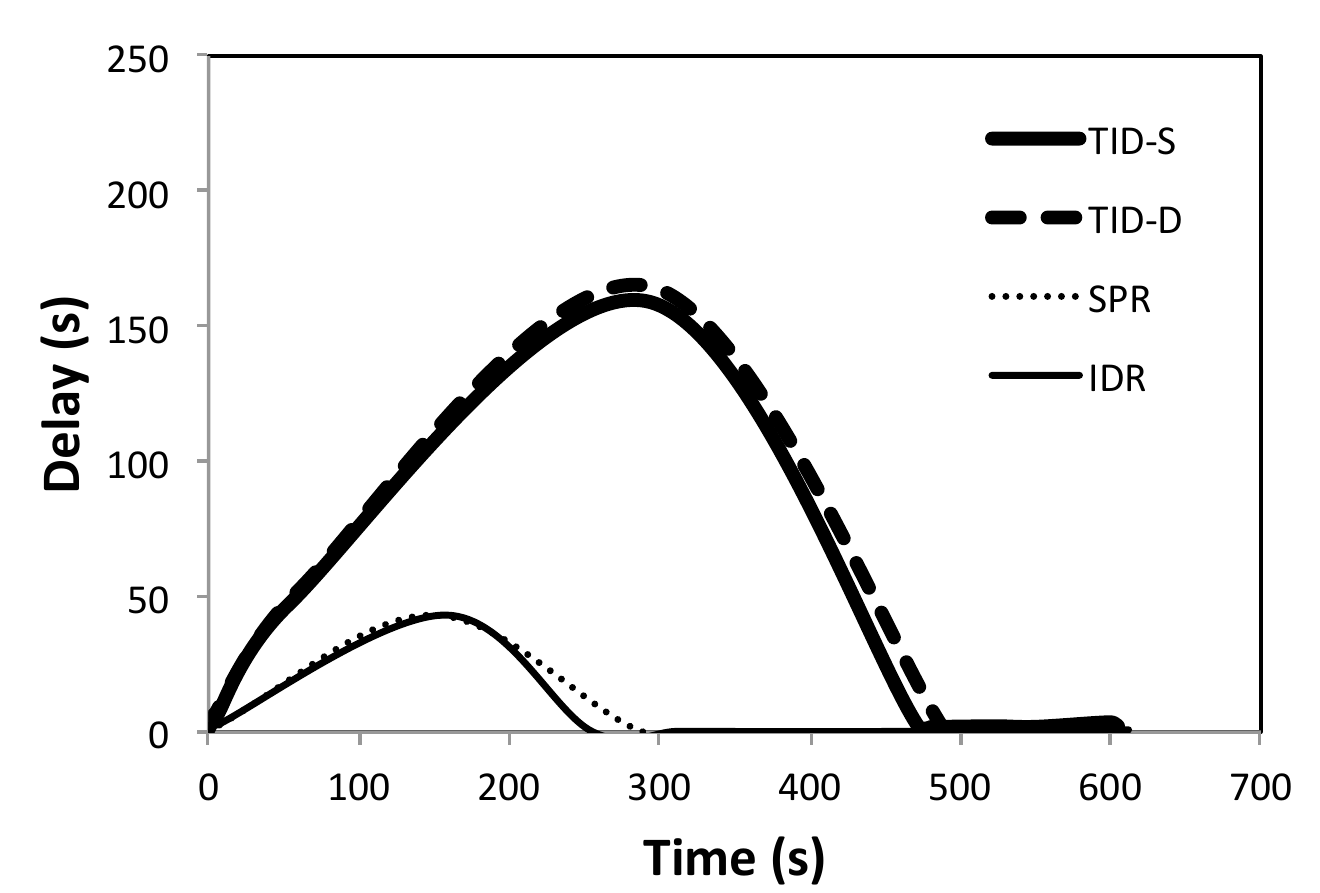}
		\subcaption{\scriptsize Fig. 19: 60K messages per minute}
	\end{subfigure}
	~
	\begin{subfigure}[b]{0.24\textwidth}
		\includegraphics[trim=0.5cm 0.5cm 0.5cm 2cm, width=0.95\textwidth]{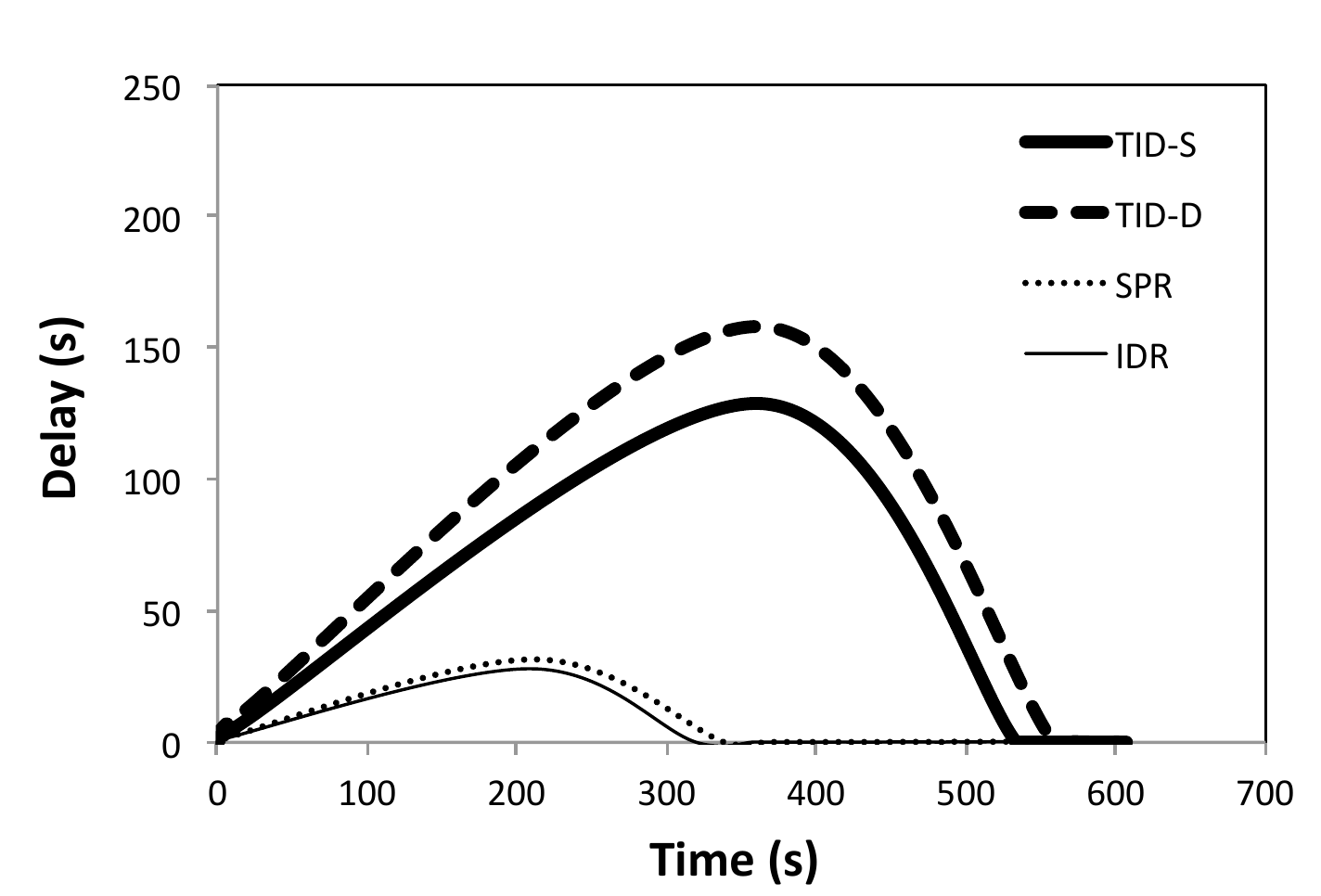}
		\caption{\scriptsize Fig. 20: 40K messages per minute}
	\end{subfigure}
\end{figure*}
\textit{Publisher Scalability: }
We increased the number of publishers from 100 to 500 to measure impact of increasing advertisements in the systems. Fig. 7 shows that the average advertisement delay in unclustered $\mathbb{S}_{e}$ (legend PADRES) was 266\% higher than the clustered $\mathbb{S}_{e}$ (legend OctopiA) and remained nearly constant. The reasons of this difference are larger lengths of advertisement--trees and extra IMs generated to detect and discard duplicates by PADRES. \texttt{OctopiA} generates advertisements--trees of length 1, and extra IMs are not required in ABP. Fig. 8 shows the average size of CLTs (in \texttt{OctopiA}) was reduced by 93\% when compared with SRTs (in PADRES). PADRES broadcasts an advertisement to all brokers in an overlay, while \texttt{OctopiA} forwards an advertisement to the secondary brokers in the host region of a publisher, which dramatically reduced size of CLTs. Fig. 9 shows that the number of IMs generated by ABP in PADRES was 88.5 times more than \texttt{OctopiA}. There are two reasons of this large difference: (i) PADRES broadcast advertisement in whole overlay, while OctopiA forwards advertisement only to brokers in the host region of the publisher, and (ii) PADRES generates a large number of extra IMs (upto 80\% of the total traffic \cite{Li_ADAP}) to detect and discard duplicates, while ABP in \texttt{OctopiA} does not generate extra IMs.\\
\textit{Subscriber Scalability:}
We increased the number of subscribers from 1000 to 10000, with 100 publishers, to study impact of increasing subscriptions. Fig. 11 shows that the average size of PRTs in PADRES was 33\% smaller than \texttt{OctopiA}. PADRES forwards subscriptions in the reverse direction of the matching advertisement--trees, while \texttt{OctopiA} saves more subscriptions due to cluster--level subscription broadcast. This is the reason that \texttt{OctopiA} generated 32.5\% more IMs than PADRES in SBP (Fig. 12). Although \texttt{OctopiA} saved more subscriptions and generated more IMs in SBP, Fig. 10 shows that subscription delay is OctopiA was 34\% less than PADRES. The reason of this difference is the less matching delay due to smaller size of CLTs than SRTs (Fig. 8).\\
\textit{Publication Delivery:} This part compares static and dynamic routing by \textit{forcing} IDR to use static routing paths. Using this approach, we computed performance penalty due to selecting dynamic routing paths in IDR. Fig. 13 indicates that the average publication delay in \texttt{OctopiA} was 25\% less than PADRES. The reason of this difference is the additional payload due to carrying TIDs with each publications onto longer routing paths generated by PADRES. There was a minimal difference in the publication delivery delay in static and dynamic routing in both cases. The collected data indicated that the publication delay by SPR algorithm was 0.8\% less than IDR algorithm, while TID--S was 0.6\% less than TID--D algorithm. This slight improvement with SPR and TID--S algorithms is due to the fact that static routing does not do additional checks to find alternative routing paths. The average matching delay in \texttt{OctopiA} was 11\% higher than PADRES (Fig. 14). PADRES performs matching at the host brokers of interested subscribers, while \texttt{OctopiA} does this at every broker in a routing path. Fig. 14 also illustrates that the difference in matching delay decreased when the number of subscribers increased, which decreased the number of  \textit{forward--only} brokers. Although \texttt{OctopiA} generated more IMs in SBP with larger size of PRTs (Figs. 11, 12), Fig. 15 shows that \texttt{OctopiA} generated 4.5\% less IMs than PADRES in publications routing. This indicates that \texttt{OctopiA} generated shorter publication routing paths in clustered $\mathbb{S}_{e}$. The difference in IMs diminished with the increase in the number of subscribers, due to decrease in \textit{forward--only} brokers.\\
\textit{Dynamic Routing:}
We now discuss dynamic routing and analyse stability of \texttt{OctopiA} when publications start queueing up due to a bursty publisher. The stability analysis demonstrate our approach by showing how quickly \texttt{OctopiA} converges to a normal state after an HRP finishes a burst. To activate dynamic routing, we used Eq. 3 in PADRES and \texttt{OctopiA} for a fair comparison. Eq. 3 has great similarity with the link utilization ratio that PADRES uses \cite{Li_ADAP}. We set the value of $\tau$ to 10 and $t_{w}$ to 50 milliseconds for 10000 subscribers and 100 publishers. Each publisher generated 1000 publications with varying rate of 100 to 120 publications per minute. We also evaluated static routing using SPR and TID--S with the same workload. To support dynamic routing, PADRES requires that a subscription should match with multiple advertisements to increase possibility of finding alternative routing paths \cite{Li_ADAP}. To handle this limitation, we increased selectivity of each subscriber from 2\% to 4\% to receive publications from 4 different publishers, however, availability of 4 alternative routing paths is not guaranteed as publishers and subscribers are anonymous. To exert more load on iLinks and aLinks, all clusters of $\mathbb{S}_{e}$ were the target clusters of the HRP with 0.1\% matching subscriptions. No subscriber (for HRP) was hosted by the host broker of the HRP. We used varying rate of the HRP to study impact of shorter and longer bursts. The HRP sent 50K publications at rates of 100K, 80K, 60K, and 40K publications per minute. We ran each simulation 20 times, for 80 times in total. The burst continued for 30 to 75 seconds depending on the burst rate. Each point in Figs. 17--20 is a maximum delivery delay of 5000 publications received in a sequence. This approach helps in analysing the stability without graphing too many points. On average, IDR algorithm stabilized \texttt{OctopiA} 35 seconds before SPR algorithm and generated only 0.04\% more IMs. In the first 157 seconds, the maximum publication delay was the same in the two algorithms and no tendency toward stability was observed. This indicates that, due to the high rates of publications, the state of the system (\texttt{OctopiA}) did not converge to normal until the condition $CE < 1$ was maintained for some time (on average 60 seconds for the four simulations). The average value of $Q_{\ell}$ of target links at the brokers in routing paths when IDR algorithm was used is 32\% less than SPR algorithm. Thanks to the bit--vector approach, which uses $CIV^i_{p}$ for dynamic routing to improve performance of \texttt{OctopiA} by adding minimum copies of publications in the congested output queues. The analysis of the collected data indicates that 51\% publications in IDR algorithm and 48\% in SPR algorithm had delay less than one second. This indicates that IDR algorithm not just stabilized the system before SPR algorithm but also decreased the average delays. Figs. 17--20 indicate that static and dynamic routing in PADRES has varied performance. In some cases, static routing performed better than dynamic (Figs. 17, 20), while in other dynamic routing performed better than static routing (Fig. 18) or both had same performance (Fig. 19). For 80 simulations, TID--S performed better than TID--D in 65\%, TID--D performed better than TID--S in 25\%, while performance was same in 10\% simulations. On average, publication delay by TID--D algorithm is 18\% higher than the TID--S algorithm, which indicates that dynamic routing by PADRES when selectivity is 4\% is poor than static routing. Average publication delay by \texttt{OctopiA} is 54\% less than PADRES, which indicates an excellent improvement in burst scenario. There are multiple reasons for better performance. First, PADRES maintains subscriptions and TIDs in separate space for algorithmic reasons. When routing switches to dynamic, after finding matching subscriptions, PADRES performs a sequential search to find TIDs of the matched subscriptions to search uncongested next destination paths. As covering is not supported for cyclic overlays, linear search over a complete subscription space is expansive and dwarf improvements of dynamic routing by increasing path selection delay. As Fig. 16 illustrates, the path selection delay in TID--D algorithm is 1230\% higher than IDR and 600\% higher than TID--S algorithms. On the other side, path selection delay in IDR algorithm is just 7\% higher than SPR algorithm. Thanks to the bit--vector approach of IDR algorithm that exploits structuredness of $\mathbb{S}_{e}$ to find next destination clusters without a linear search over saved subscriptions. Second, PADRES does not assure dynamic routing as overlapping advertisement--trees for a subscription are not guaranteed even if multiple copies of the same subscription is recorded at brokers \cite{Li_ADAP}. Further, PADRES does not guarantee of having multiple brokers with overlapping advertisement--trees in a dynamic routing path. One broker may have overlapping advertisement--trees for a subscription to support dynamic routing, while other brokers, in the same dynamic routing path, may not support dynamic routing due to the unavailability of overlapping advertisement--trees. As a result, static routing performed better than dynamic routing in most of the cases. \texttt{OctopiA} uses structuredness of SCOT to avoid these limitations in offering inter--cluster dynamic routing. Third, longer routing paths generated by PADRES in unclustered $\mathbb{S}_{e}$ causes higher delivery delays. Fourth, payload of a publication in PADRES is higher due to carrying TIDs. On average, TID--S stabilized PADRES almost 15 seconds before TID--D routing algorithm, while SPR and IDR algorithms stabilized \texttt{OctopiA} 200 seconds before PADRES.
\section{Related Work}
Although content--based routing in broker--based pub/sub systems has been focus of many research efforts \cite{JEDI,Rebeca,Kyra,MEDYN}, only SIENA \cite{SIENA_WIDE_AREA} and PADRES \cite{PADRESBookChapte} support routing in cyclic overlays. SIENA uses latency--based distance--vector algorithms for generating routing paths \cite{carz_thesis}. The system, however, does not generate advertisement--trees of shortest lengths and dynamic routing is not supported. Shuping et al \cite{MERC} divides a cyclic overlay into clusters to apply content--based inter-- and intra--cluster routing of publications. The approach embeds a list of destination brokers to deliver publications, which consumes network bandwidth inefficiently. A broker has to be aware of other brokers in a cluster, which increases topology maintenance cost. Dynamic routing requires updates in routing tables, which increases network traffic and delivery delays. Li et al \cite{Li_ADAP} offers dynamic routing without making updates in routing tables and their approach is closet to our work. Each advertisement--tree is uniquely identified and a large number of IMs are generated in advertisement broadcast to detected duplicates. This technique relies on intersecting advertisement--trees and does not support dynamic routing for subscriptions matching with single advertisement. Intersecting advertisement--trees are not always possible even if a subscription matches with multiple advertisements. \texttt{OctopiA} neither generates redundant IMs in advertisement broadcast nor requires carrying unique identification to discard duplicates. Resultantly, publications do not carry extra payload to identify routing paths to avoid message loops. Thanks to the structuredness of SCOT overlay, which inherently support subgrouping and eliminates tight coupling. \texttt{OctopiA} requires a minimal topology maintenance cost as SCOT brokers are decoupled and aware of their direct neighbours only. Our algorithms generate advertisement--trees of length 1 and offers inter--cluster dynamic routing without requiring updates in routing tables. All these properties make \texttt{OctopiA} scalable and suitable for large pub/sub systems.
\section{Conclusion}
This paper introduces the first pub/sub system, \texttt{OctopiA}, that offers inter--cluster dynamic routing of publications. Tight coupling in pub/sub systems is one of the main obstacle that requires updates in routing tables to generate alternative routing paths. \texttt{OctopiA} uses a novel concept of subgrouping, which divides subscriptions into exclusive disjoint subsets. Subgrouping is realized using the structuredness of a clustered SCOT overlay, where subscriptions in a subgroup are available in an exclusive host cluster through the cluster--level subscription broadcast. An advertisement--trees in clustered SCOT has length 1. These patterns of subscriptions and advertisements broadcast eliminate tight coupling to enable inter--cluster dynamic routing without making updates in routing tables. As subscriptions are available only in their host clusters, \texttt{OctopiA} uses a bit--vector to identify TSCs to avoid false positives in inter--cluster routing of publications. \texttt{OctopiA} does not generate extra IMs to detect and discard duplicate advertisements and requires no path identification to route publications. Evaluation with real word data indicates that SPR and and IDR algorithms scale well when compared with state--of--the--art TID--based static and dynamic routing. Support for intra--cluster dynamic routing is part of the future work.
\bibliographystyle{plain}
\bibliography{refs_p2}

\begin{thebibliography}{10}

\bibitem{QoS_Survey}
P.~Bellavista, A.~Corradi, and A.~Reale.
\newblock Quality of service in wide scale publish-subscribe systems.
\newblock {\em IEEE Communications Surveys Tutorials}, 16(3):1591--1616, Third
  2014.

\bibitem{MMOG_Canas}
C{\'e}sar Ca\~{n}as, Kaiwen Zhang, Bettina Kemme, J\"{o}rg Kienzle, and
  Hans-Arno Jacobsen.
\newblock Publish/subscribe network designs for multiplayer games.
\newblock In {\em Proceedings of the 15th International Middleware Conference},
  Middleware '14, pages 241--252, New York, NY, USA, 2014. ACM.

\bibitem{Kyra}
Fengyun Cao and J.P. Singh.
\newblock Efficient event routing in content-based publish-subscribe service
  networks.
\newblock In {\em INFOCOM 2004. Twenty-third AnnualJoint Conference of the IEEE
  Computer and Communications Societies}, volume~2, pages 929--940 vol.2, March
  2004.

\bibitem{MEDYN}
Fengyun Cao and J.P. Singh.
\newblock Medym: match-early and dynamic multicast for content-based
  publish-subscribe service networks.
\newblock In {\em Distributed Computing Systems Workshops, 2005. 25th IEEE
  International Conference on}, pages 370--376, June 2005.

\bibitem{carz_thesis}
Antonio Carzaniga.
\newblock Architectures for an event notification service scalable to wide-area
  networks, phd thesis, politecnico di milano.
\newblock 1998.

\bibitem{SIENA_WIDE_AREA}
Antonio Carzaniga, David~S. Rosenblum, and Alexander~L. Wolf.
\newblock Design and evaluation of a wide-area event notification service.
\newblock {\em ACM Trans. Comput. Syst.}, 19(3):332--383, August 2001.

\bibitem{Congestion}
Mingwen Chen, Songlin Hu, Vinod Muthusamy, and Hans-Arno Jacobsen.
\newblock {Congestion Avoidance with Incremental Filter Aggregation in
  Content-Based Routing Networks}.
\newblock In {\em IEEE 35th International Conference on Distributed Computing
  Systems (ICDCS)}, June 2015.

\bibitem{JEDI}
G.~Cugola, Elisbetta Di~Nitto, and A.~Fuggetta.
\newblock The jedi event-baased infrastructure and its application to the
  development of the opss wfms.
\newblock {\em Software Engineering, IEEE Transactions on}, 27(9):827--850,
  2001.

\bibitem{PNUTS}
Cooper et~al.
\newblock Pnuts: Yahoo!'s hosted data serving platform.
\newblock {\em Proc. VLDB Endow.}, 1(2):1277--1288, August 2008.

\bibitem{PADRESBookChapte}
Hans-Arno et~al.
\newblock {The PADRES Publish/Subscribe System}.
\newblock In {\em Principles and Applications of Distributed Event-Based
  Systems}, pages 164--205. IGI Global, 2010.

\bibitem{agg15}
Navneet et~al.
\newblock {Minimizing the Communication Cost of Aggregation in
  Publish/Subscribe Systems}.
\newblock In {\em 35th IEEE International Conference on Distributed Computing
  Systems (ICDCS)}, 2015.

\bibitem{WormHole}
Yogeshwer~Sharma et~al.
\newblock Wormhole: Reliable pub-sub to support geo-replicated internet
  services.
\newblock In {\em NSDI 15}, pages 351--366, Oakland, CA, May 2015. USENIX
  Association.

\bibitem{MANY_FACES}
Patrick~Th. Eugster, Pascal~A. Felber, Rachid Guerraoui, and Anne-Marie
  Kermarrec.
\newblock The many faces of publish/subscribe.
\newblock {\em ACM Comput. Surv.}, 35(2):114--131, June 2003.

\bibitem{CPUG_Book}
Richard Hammack, Wilfried Imrich, and Sandi Klavzar.
\newblock {\em Handbook of Product Graphs}.
\newblock CRC Press, Inc., 2nd edition, 2011.

\bibitem{MS_PULSE}
Microsoft Bing~Pulse (http://pulse.bing.com/).

\bibitem{G_CLOUD_PS}
Google Cloud~Pub/Sub (https://cloud.google.com/pubsub/).

\bibitem{MERC}
Shuping Ji, Chunyang Ye, Jun Wei, and Hans-Arno Jacobsen.
\newblock Merc: Match at edge and route intra-cluster for content-based
  publish/subscribe systems.
\newblock In {\em Proceedings of the 16th Annual Middleware Conference}, pages
  12--24, 2015.

\bibitem{reza-thesis}
Reza~Sherafat Kazemzadeh.
\newblock {\em {Overlay Neighborhoods for Distributed Publish/Subscribe
  Systems}}.
\newblock PhD thesis, University of Toronto, 2012.

\bibitem{reza-partition}
Reza~Sherafat Kazemzadeh and Hans-Arno Jacobsen.
\newblock {Partition-tolerant Distributed Publish/Subscribe Systems}.
\newblock In {\em 30th IEEE Symposium on Reliable Distributed Systems (SRDS
  2011)}, pages 101--110, Madrid, Spain, October 2011. IEEE.

\bibitem{reza_bulk}
R.S. Kazemzadeh and H.-A. Jacobsen.
\newblock Publiy+: A peer-assisted publish/subscribe service for timely
  dissemination of bulk content.
\newblock In {\em Distributed Computing Systems (ICDCS), 2012 IEEE 32nd
  International Conference on}, pages 345--354, June 2012.

\bibitem{KAFKA}
Jay Kreps, Neha Narkhed, and Jun Rao.
\newblock Kafka: a distributed messaging system for log processin.
\newblock {\em NetDB Workshop}, pages 01--07, 2011.

\bibitem{Li_ADAP}
Guoli Li, Vinod Muthusamy, and Hans-Arno Jacobsen.
\newblock Adaptive content-based routing in general overlay topologies.
\newblock Middleware '08, pages 1--21, New York, NY, USA, 2008. Springer-Verlag
  New York, Inc.

\bibitem{Com_clustering}
Wei Li, Songlin Hu, Jintao Li, and H.-A. Jacobsen.
\newblock Community clustering for distributed publish/subscribe systems.
\newblock In {\em Cluster Computing (CLUSTER), 2012 IEEE International
  Conference on}, pages 81--89, Sept 2012.

\bibitem{work_flow}
H.A. Jacobsen R.~Hull M.~Sadoghi, M.~Jergler and R.~Vaculin.
\newblock Safe distribution and parallel execution of data-centric workflows
  over publish/subscribe abstraction.
\newblock {\em IEEE Transaction on Knowledge and Data Engineering},
  27(10):2824--2838, 2015.

\bibitem{routing_book}
Deepankar Medhi and Karthikeyan Ramasamy.
\newblock {\em Network Routing: Algorithms, Protocols, and Architectures}.
\newblock Morgan Kaufmann Publishers Inc., San Francisco, CA, USA, 2007.

\bibitem{muthy_thesis}
Vinod Muthusamy.
\newblock {\em {Flexible Distributed Business Process Management}}.
\newblock PhD thesis, University of Toronto, 2011.

\bibitem{theGeneralizedPUG}
E.~Parsonage, H.X. Nguyen, R.~Bowden, S.~Knight, N.~Falkner, and M.~Roughan.
\newblock Generalized graph products for network design and analysis.
\newblock In {\em Network Protocols (ICNP), 2011 19th IEEE International
  Conference on}, pages 79--88, Oct 2011.

\bibitem{Rebeca}
Helge Parzyjegla, Daniel Graff, Arnd Schr{\"o}ter, Jan Richling, and Gero
  M{\"u}hl.
\newblock {\em From Active Data Management to Event-Based Systems and More:
  Papers in Honor of Alejandro Buchmann on the Occasion of His 60th Birthday},
  chapter Design and Implementation of the Rebeca Publish/Subscribe Middleware,
  pages 124--140.
\newblock Springer Berlin Heidelberg, Berlin, Heidelberg, 2010.

\bibitem{algo_trading}
Mohammad Sadoghi, Martin Labrecque, Harsh Singh, Warren Shum, and Hans-Arno
  Jacobsen.
\newblock Efficient event processing through reconfigurable hardware for
  algorithmic trading.
\newblock {\em Proc. VLDB Endow.}, 3(1-2):1525--1528, September 2010.

\bibitem{TariqPLEROMA}
Muhammad~Adnan Tariq, Boris Koldehofe, Sukanya Bhowmik, and Kurt Rothermel.
\newblock Pleroma: A sdn-based high performance publish/subscribe middleware.
\newblock Middleware '14, pages 217--228, 2014.

\end{thebibliography}
\end{document}